\documentclass[10pt,sigconf]{acmart}
\usepackage{graphicx}
\usepackage{booktabs} 
\usepackage{subfigure}
\usepackage[]{amsmath,amsfonts,amssymb}
\usepackage{breqn} 
\usepackage{mathtools}
\usepackage{siunitx}
\usepackage{epstopdf}
\usepackage{multirow}
\usepackage{xcolor}
\usepackage{enumitem}
\usepackage{draftwatermark}
\SetWatermarkText{Camera Ready}
\SetWatermarkScale{0.5}
\DeclareMathOperator*{\argmax}{arg\,max}
\copyrightyear{2018}
\acmYear{2018}
\setcopyright{acmcopyright}
\acmConference[mmNets'18]{2nd ACM Workshop on Millimeter Wave Networks and Sensing Systems}{October 29, 2018}{New Delhi, India}
\acmBooktitle{2nd ACM Workshop on Millimeter Wave Networks and Sensing Systems (mmNets '18), October 29, 2018, New Delhi, India}
\acmPrice{15.00}
\acmDOI{10.1145/3264492.3264503}
\acmISBN{978-1-4503-5928-3/18/10}
\begin{document}
	\title{Direction of Arrival and Center Frequency Estimation for Impulse Radio Millimeter Wave Communications}
	\vspace{-20mm}
	\author{Shree Prasad M., Trilochan Panigrahi}
	\affiliation{%
		\institution{National Institute of Technology Goa, India}
		\postcode{403401}
	}
	\email{shreeprasad,tpanigrahi@nitgoa.ac.in}
	\vspace{-20mm}
	\author{Mahbub Hassan}
	\affiliation{%
		\institution{University of New South Wales, Sydney, Australia}
	}
	\email{mahbub@cse.unsw.edu.au}
	\vspace{-5mm}
	\begin{abstract}
		\vspace{-1mm}
		The 30-300GHz millimeter wave (mmWave) band is currently being pursued to combat the rising capacity demands in 5G, WiFi, and IoT networks. Due to the high frequency, impulse radio (IR) in this band is better suited for positioning than other existing low-frequency bands. Besides precision positioning, the exceptionally wide bandwidth also enables concurrent use of multiple center frequencies in the same application, which opens up additional avenues of information encoding in IR mmWave networks. In this paper, we propose a new mmWave IR framework that can simultaneously detect direction of arrival (DOA) as well as the center frequency of the transmitted pulse. Based on the emerging graphene-based transceivers, we evaluate the performance of the proposed framework in the higher frequency region of mmWave band (100-300GHz). Numerical experiments demonstrate that the proposed framework can detect the DOA of a 0.1 $\mu$W mmWave pulse within 1 degree of precision at 20 meters, and classify three different center frequencies with 100\% accuracy from a distance of 10 meters. These performances could be further improved by trading off the pulse rate of the system.
		\vspace{-4mm}
	\end{abstract}
	\vspace{-6mm}
	\vspace{-7mm}
	\maketitle
	\vspace{-4mm}
	\section{Introduction}
	To combat the severe spectrum shortage in current communication bands, the industry is exploring the use of 30-300GHz mmWave band in both 5G and WiFi networks. While most mmWave developments have focussed on conventional \textit{continuous wave} transmissions, researchers are beginning to explore the \textit{impulse radio} (IR) benefits of this band \cite{5G_TOA2}. Impulse refers to an extremely short pulse on the order of nano or pico seconds with its energy distributed over a wide bandwidth. The high frequency enables mmWave pulses to localize objects very precisely. Indeed, many new positioning applications, such as precise vehicle positioning for driverless cars \cite{5G_VEH_POS, 5G_TOA2}, are emerging based on mmWave IR. 
	
	We explore other opportunities of IR mmWave beyond positioning. We observe that the extremely wide bandwidth of mmWave band opens up new opportunities to design multiple non-overlapping pulses with different center frequencies contained within the same communication band. It is thus feasible to conceive new IR mmWave applications that can code information in the center frequency of a pulse. For example, energy constrained Internet of Things could simply send a single pulse to update their status coded in its center frequency, provided the base station could detect the center frequency of the received pulse. When combined with positioning capability of IR, the same single pulse could provide both localization and status information for the IoT at minimal power consumption.\\     	
	\begin{figure*}[htb]
		\subfigure{
			\includegraphics[width=0.425\columnwidth,height=2cm]{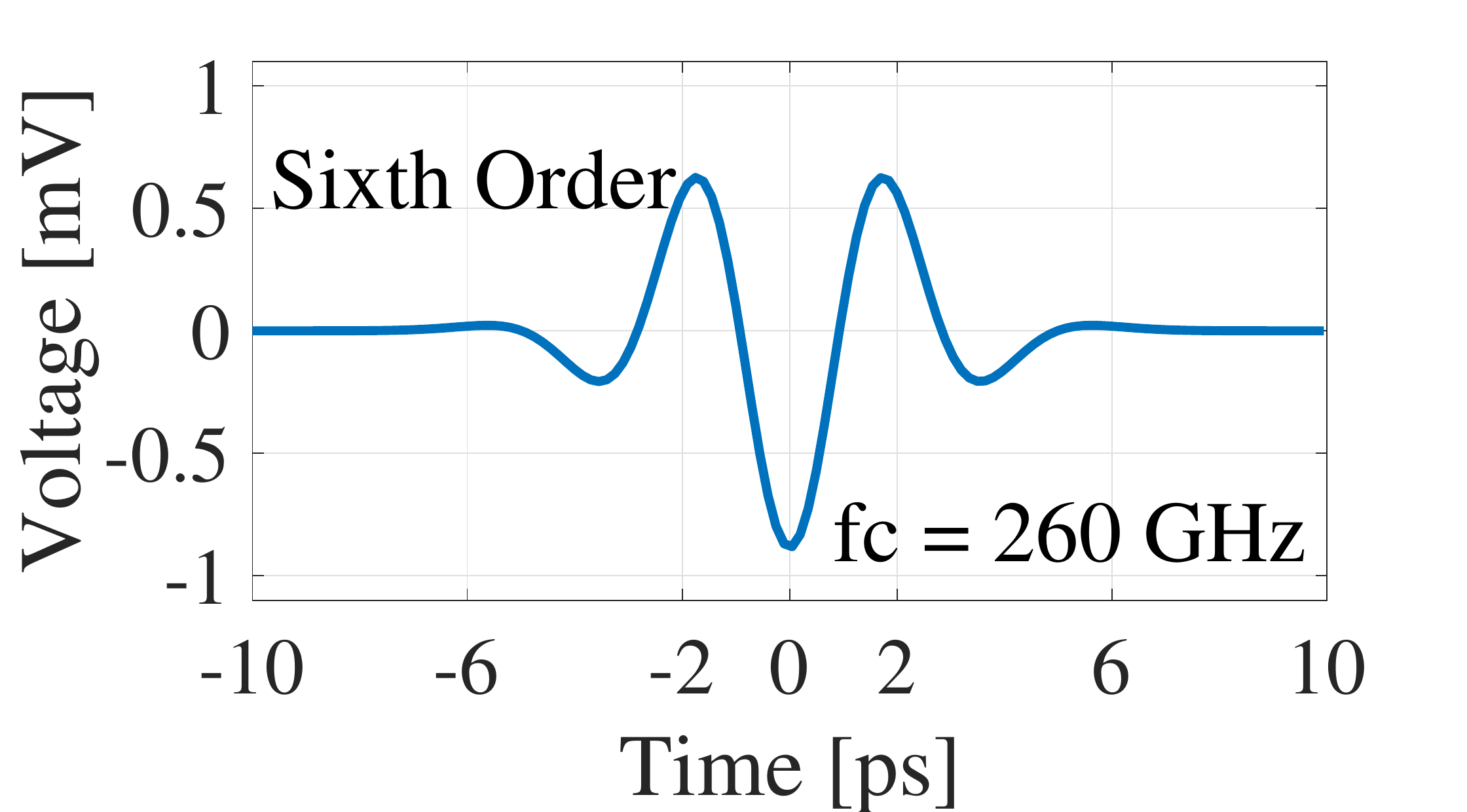}
		}
		\hspace{-5mm}
		\subfigure{
			\includegraphics[width=0.425\columnwidth,height=2cm]{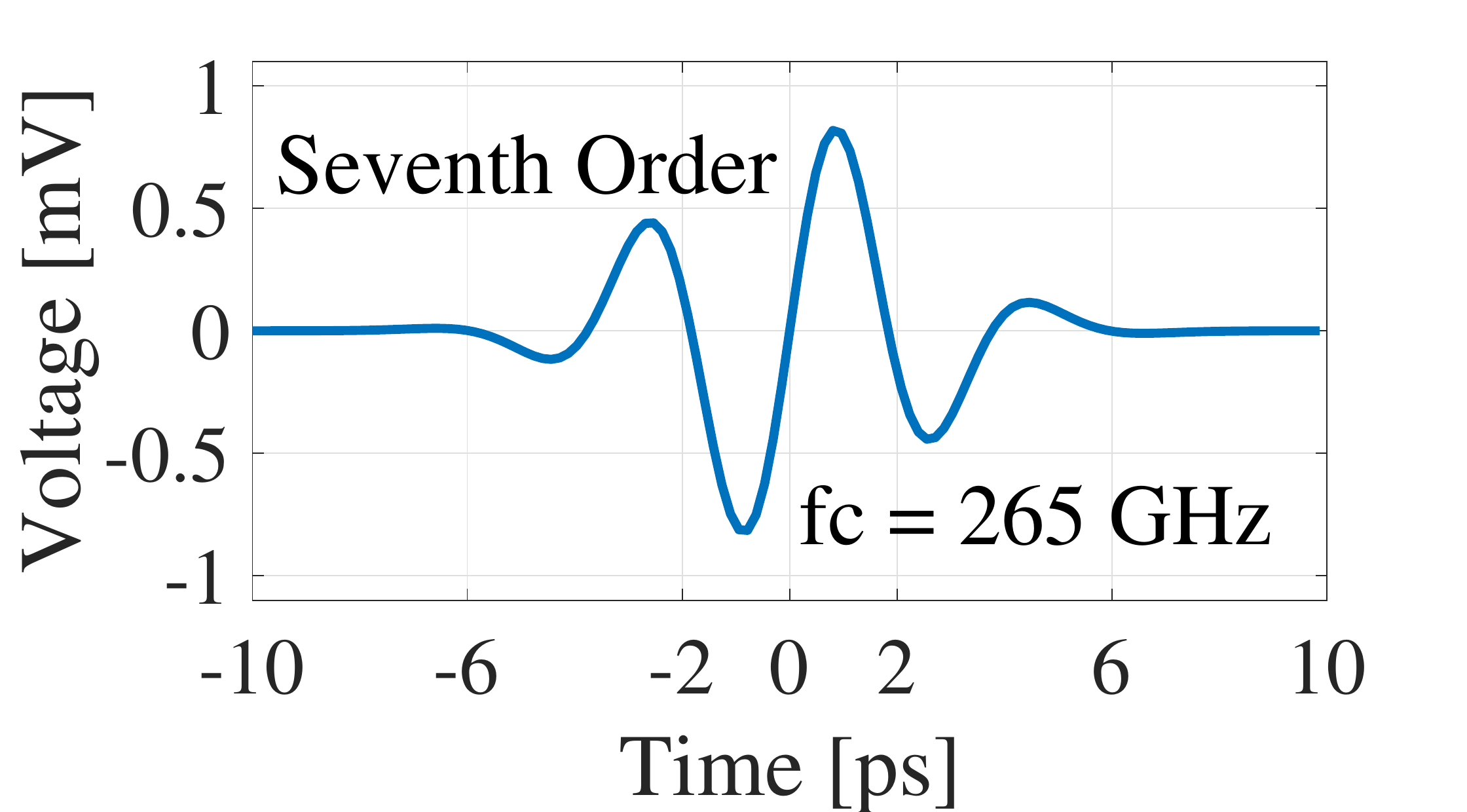}	
		}
		\hspace{-5mm}
		\subfigure{
			\includegraphics[width=0.425\columnwidth,height=2cm]{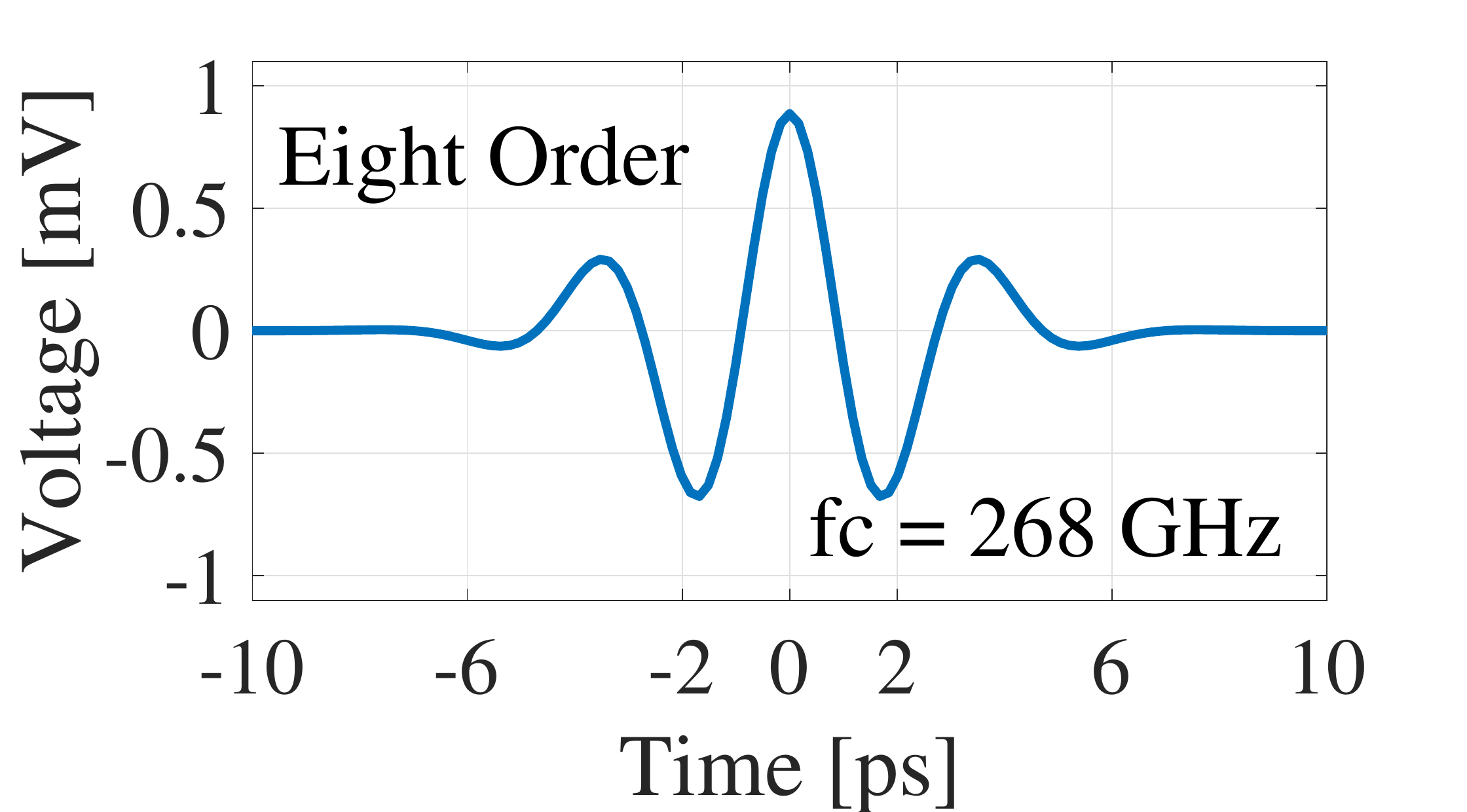}	
		}
		\hspace{-5mm}
		\subfigure{
			\includegraphics[width=0.425\columnwidth,height=2cm]{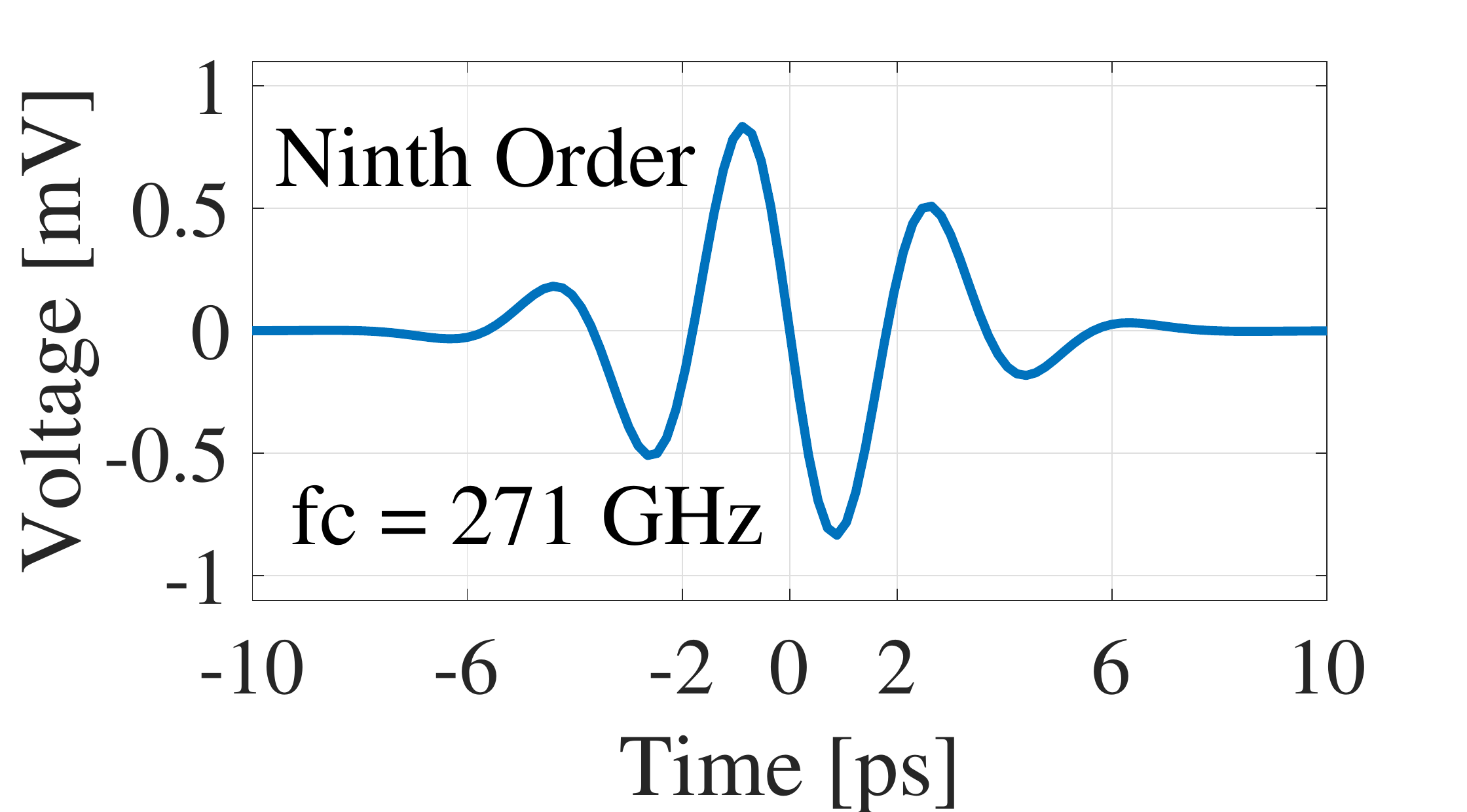}	
		}
		\hspace{-5mm}
		\subfigure{
			\includegraphics[width=0.425\columnwidth,height=2cm]{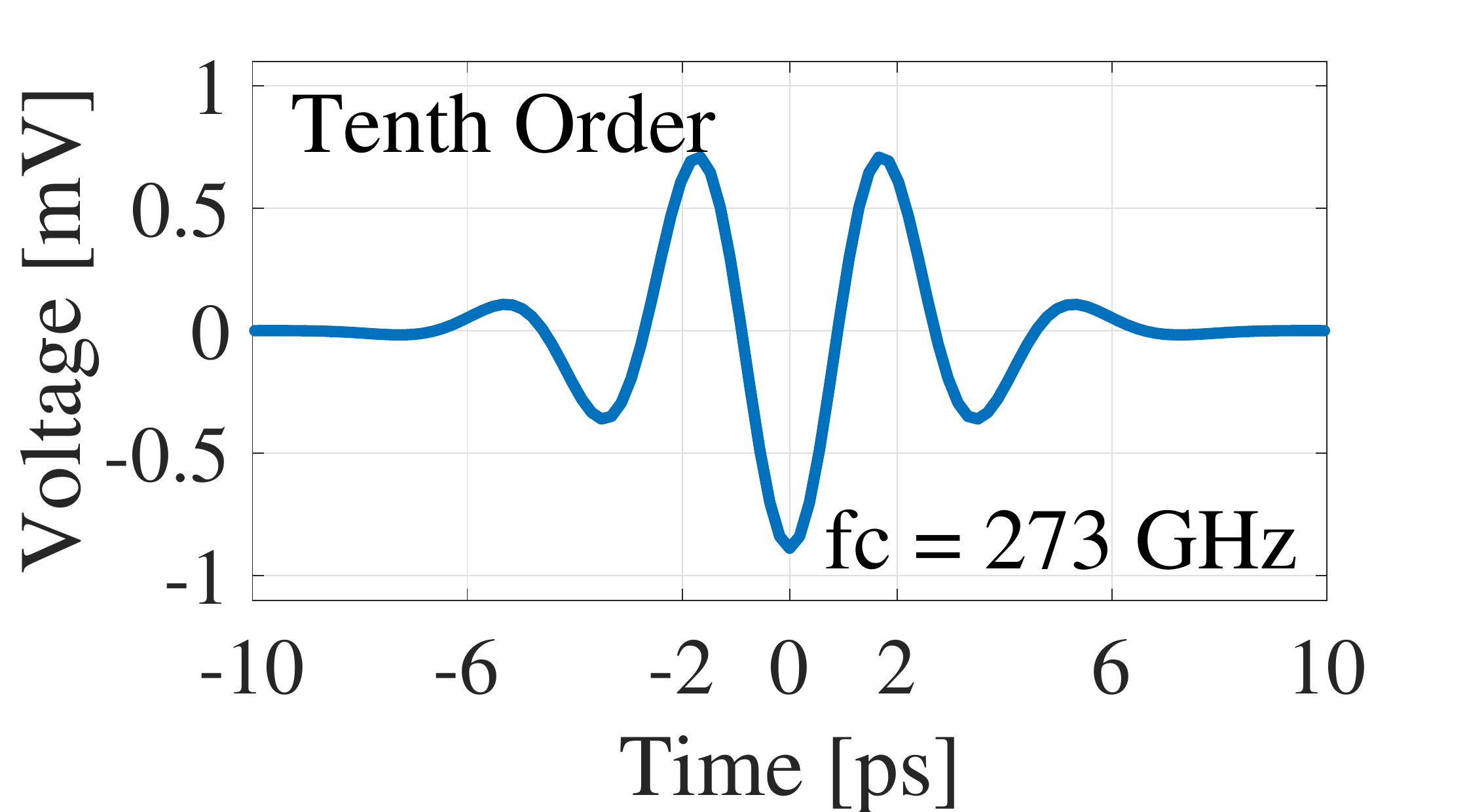}	
		}
		\vspace{-6mm}
		\caption{Higher order Gaussian pulses of 0.1 \si{\micro}W power with different center frequencies
		}
		\vspace{-4mm}
		\label{fig:D_PLT}
	\end{figure*}
	\begin{table*}[h]
		\centering
		\caption{Half power bandwidth of higher order Gaussian pulses at different center frequencies}
		\vspace{-3.25mm}
		\label{tabl:Table1}
		\resizebox{0.95\textwidth}{0.125\textwidth}{
			\begin{tabular}{|c|c|c|c|c|c|c|c|c|c|c|c|c|c|}
				\hline
				$\boldsymbol{n}$                  & $\boldsymbol{a_{n}}$                  & $\boldsymbol{f_{c}}$ [GHz]                  & $\boldsymbol{f_{l}}$ [GHz]                     &$\boldsymbol{f_{h}}$ [GHz]                      & $\boldsymbol{B_{3\:dB}}$ [GHz]                     & $\boldsymbol{T_{g}}$ [ps]                    & $\boldsymbol{n}$                 & $\boldsymbol{a_{n}}$                 &$\boldsymbol{f_{c}}$ [GHz]     & $\boldsymbol{f_{l}}$ [GHz]    & $\boldsymbol{f_{h}}$ [GHz]     & $\boldsymbol{B_{3\:dB}}$ [GHz]        & $\boldsymbol{T_{g}}$ [ps]   \\ \hline
				\multirow{3}{*}{\textbf{1}} & \multirow{3}{*}{\resizebox{0.3\columnwidth}{0.07\columnwidth}{$\sqrt{4\sqrt{\pi}\sigma^{3} E_{g}}$}} & \multirow{3}{*}{200} & \multirow{3}{*}{96.32} & \multirow{3}{*}{327.31} & \multirow{3}{*}{230.98} & \multirow{3}{*}{7.95} & \multirow{3}{*}{\textbf{6}}  & \multirow{3}{*}{\resizebox{0.3\columnwidth}{0.1\columnwidth}{$\sqrt{\frac{E_{g}\sqrt{\pi}\sigma^{13}}{81.2109}}$}}  & \color{blue}\textbf{131}   & 100.883 & 163.64 & 62.76   & 29.75 \\ \cline{10-14} 
				&                     &                      &                        &                         &                         &                       &                     &                      & 200   & 154.01  & 249.84 & 95.82   & 19.49 \\ \cline{10-14} 
				&                     &                      &                        &                         &                         &                       &                     &                      & \color{blue}\textbf{260}   & 200.22  & 324.80 & 124.57  & 14.99 \\ \hline
				\multirow{3}{*}{\textbf{2}} & \multirow{3}{*}{\resizebox{0.3\columnwidth}{0.1\columnwidth}{$\sqrt{\frac{E_{g}\sqrt{\pi}\sigma^{5}}{\left ( 3/8 \right )}}$}} & 150                  & 92.54                  & 216.22                  & 123.68                  & 15.00                 & \multirow{3}{*}{\textbf{7}}  & \multirow{3}{*}{\resizebox{0.3\columnwidth}{0.1\columnwidth}{$\sqrt{\frac{E_{g}\sqrt{\pi}\sigma^{15}}{527.8711}}$}}  & \color{blue}\textbf{128}   & 100.65  & 157.46 & 56.80   & 32.89 \\ \cline{3-7} \cline{10-14} 
				&                     & 200                  & 123.38                 & 288.30                  & 164.91                  & 11.23                 &                     &                      & 210   & 165.14  & 258.33 & 93.19   & 20.05 \\ \cline{3-7} \cline{10-14} 
				&                     & 250                  & 154.23                 & 360.37                  & 206.14                  & 9.00                  &                     &                      & \color{blue}\textbf{265}   & 208.39  & 325.99 & 117.60  & 15.88 \\ \hline
				\multirow{3}{*}{\textbf{3}} & \multirow{3}{*}{\resizebox{0.3\columnwidth}{0.1\columnwidth}{$\sqrt{\frac{E_{g}\sqrt{\pi}\sigma^{7}}{\left ( 15/16 \right )}}$}} & 148                  & 100.89                 & 200.85                  & 99.96                   & 18.62                 & \multirow{3}{*}{\textbf{8}}  & \multirow{3}{*}{\resizebox{0.3\columnwidth}{0.1\columnwidth}{$\sqrt{\frac{E_{g}\sqrt{\pi}\sigma^{17}}{3959.0332}}$}}  & \color{blue}\textbf{125}   & 99.95   & 151.86 & 51.90   & 32.89 \\ \cline{3-7} \cline{10-14} 
				&                     & 200                  & 136.33                 & 271.42                  & 135.08                  & 13.78                 &                     &                      & 190   & 151.92  & 230.82 & 78.90   & 23.69 \\ \cline{3-7} \cline{10-14} 
				&                     & 250                  & 170.42                 & 339.28                  & 168.85                  & 11.02                 &                     &                      & \color{blue}\textbf{268}   & 214.29  & 325.58 & 111.294 & 16.98 \\ \hline
				\multirow{3}{*}{\textbf{4}} & \multirow{3}{*}{\resizebox{0.3\columnwidth}{0.1\columnwidth}{$\sqrt{\frac{E_{g}\sqrt{\pi}\sigma^{9}}{3.28125}}$}} & \color{blue}\textbf{139}                  & 100.30                 & 181.73                  & 81.43                   & 22.89                 & \multirow{3}{*}{\textbf{9}}  & \multirow{3}{*}{\resizebox{0.3\columnwidth}{0.1\columnwidth}{$\sqrt{\frac{E_{g}\sqrt{\pi}\sigma^{19}}{33651.78223}}$}}  & \color{blue}\textbf{123}   & 99.70   & 147.87 & 48.17   & 38.81 \\ \cline{3-7} \cline{10-14} 
				&                     & 200                  & 144.31                 & 261.49                  & 117.17                  & 15.91                 &                     &                      & \color{blue}\textbf{182.5} & 147.93  & 219.41 & 71.47   & 26.16 \\ \cline{3-7} \cline{10-14} 
				&                     &\color{blue}\textbf{250}                  & 180.39                 & 326.86                  & 149.46                  & 12.73                 &                     &                      & \color{blue}\textbf{271}   & 219.67  & 325.81 & 106.132 & 17.61 \\ \hline
				\multirow{3}{*}{\textbf{5}} & \multirow{3}{*}{\resizebox{0.3\columnwidth}{0.1\columnwidth}{$\sqrt{\frac{E_{g}\sqrt{\pi}\sigma^{11}}{14.765625}}$}} & \color{blue}\textbf{134}                  & 100.41                 & 170.69                  & 70.28                   & 26.55                 & \multirow{3}{*}{\textbf{10}} & \multirow{3}{*}{	\resizebox{0.3\columnwidth}{0.1\columnwidth}{$\sqrt{\frac{E_{g}\sqrt{\pi}\sigma^{21}}{319691.9069}}$}} & \color{blue}\textbf{123}   & 100.858 & 146.56 & 45.70   & 40.91 \\ \cline{3-7} \cline{10-14} 
				&                     & 200                  & 149.86                 & 254.77                  & 104.90                  & 17.79                 &                     &                      & \color{blue}\textbf{180}   & 147.59  & 214.48 & 66.89   & 27.96 \\ \cline{3-7} \cline{10-14} 
				&                     &\color{blue}\textbf{255}                  & 191.08                 & 324.83                  & 133.75                  & 13.95                 &                     &                      & \color{blue}\textbf{273}   & 223.85  & 325.30 & 101.45  &  18.43 \\ \hline
		\end{tabular}}
		\vspace{-4mm}
	\end{table*}
	In this paper, we propose a new mmWave IR framework that can simultaneously detect DOA as well as the center frequency of the transmitted pulse. Our specific contributions are as follows:\\
		$\bullet$ We propose a method to detect both DOA and center frequency of a mmWave pulse received at a linear uniform array (ULA) antenna. While we use the well-known MUSIC algorithm for DOA estimation, we propose a new method to estimate and classify the center frequency using the concept of \textit{spectral centroid}. Although spectral centroid has been used in other applications, such as cognitive load classifications \cite{SPEECH},  to our knowledge, this is the first attempt to use it for central frequency detection in IR communications.\\
		$\bullet$ Based on the propagation and noise properties of emerging graphene-based transceivers, we analyze the performance of the proposed framework in the higher frequency region of the mmWave band (100-300GHz), which happens to be within the resonance frequency band of graphene. Further, Graphene enables the fabrication of miniaturized and reconfigurable transceivers, which is not possible with conventional materials. Numerical experiments demonstrate that the proposed framework can detect DOA of a 0.1 $\mu$W higher time derivative Gaussian pulse within 1 degree of precision at 20 meters, and classify three different center frequencies with 100\% accuracy from a distance of 10 meters. Our framework allows to improve these performances by trading off the pulse rate (e.g., IoT status update rate) of the system.
	
	The remainder of the paper is organized as follows. The higher time derivative Gaussian IR waveforms are examined in Section II followed by the channel characteristics of graphene-based transceivers in Section III.  The system model along with DOA and center frequency estimation methods are described in Section IV. We present numerical experiments in Section V and conclude the paper in Section VI.
	\vspace{-6.5mm}
	\section{Higher Time Order Derivative Gaussian Pulses}
	\vspace{-0.5mm}
	mmWave devices based on Graphene nano antennas are expected to communicate by transmitting higher order Gaussian pulses of few \si{\micro}W per pulse \cite{FEMSEC}. The time domain representation of standard Gaussian pulse $g\left( t\right) $ is given as 
	\begin{equation}\label{eqn:STD_PL}
	\vspace{-1mm}
	g\left( t\right) = \frac{1}{\sqrt{2\pi}\sigma}e^{-\frac{t^{2}}{2\sigma^2}} 
\vspace{-1mm}
	\end{equation}
	where $\sigma$ is the standard deviation of the Gaussian pulse in seconds.
	The higher time order Gaussian pulse $g_{n}\left( t\right)$  is obtained by taking time derivative of $g\left( t\right)$ and is given as 
	\vspace{-0.5mm}
	\begin{equation}\label{eqn:HTOD}
	\vspace{-1mm}
	g_{n}\left(t \right) = a_{n}\frac{d^{n}}{dt^{n}}\left( g\left( t\right)\right) 
	\vspace{-0.5mm}
	\end{equation}
	where $n$ represents the order of Gaussian pulse. The scaling factor $a_{n}$, adjusts the energy content $E_{g}$ of the pulse and is given as
	\vspace{-1mm}
	\begin{equation}
	\resizebox{!}{0.06\columnwidth}{$a_{n} =  \sqrt{\dfrac{E_{g}}{\int\limits_{-\infty}^{\infty}\left| g_{n}\left(t \right)\right|^{2} dt}}$}
	\end{equation} 
	Since Gaussian pulses are of infinite duration, the total time duration $T_{g}$ of these pulses is assumed as $10\sigma$ \cite{FMFRE}. Here, $T_{g}$ is defined as the time interval which contains more than 99.99\% of pulse energy. Hence, the power $P$ of the transmitted higher order Gaussian pulse is given as $P\; =\; E_{g}/ T_{g}$. The Fourier representation $G_{n}\left( f\right) $ of higher order Gaussian pulse is also Gaussian and is represented as 
	\vspace{-1mm}
	\begin{equation}\label{eqn:PL_FD}
	\vspace{-1mm}
	G_{n}\left( f\right) = a_{n} \left( j2\pi f\right)^{n}e^{-0.5\left( 2\pi \sigma f\right)^{2} } 
	\vspace{-1mm}
	\end{equation}
	The center frequency $f_{c}$ of the higher order Gaussian pulses is given as \cite{FMFRE}
	\begin{equation}\label{eqn:cent_freq}
	\resizebox{0.175\columnwidth}{0.05\columnwidth}{$f_{c} = \frac{\sqrt{n}}{2\pi\sigma}$}
	\end{equation}
	Table \ref{tabl:Table1} shows different center frequencies $f_{c}$ in mmWave band along with their half power frequencies $f_{l}$ and $f_{h}$ for higher order Gaussian pulses from order 1 to 10. Also shown in Table \ref{tabl:Table1} is the energy scaling factor $a_{n}$, half power bandwidth $B_{3\:dB}$ and $T_{g}$. The center frequencies of Gaussian pulses for different orders are not same as its value is adjusted to obtain nonoverlapping half power bandwidth. Further, it is observed that the number of center frequencies with nonoverlapping half power bandwidth increases with increase in the order of Gaussian pulse (in blue color). The time domain representation of higher order Gaussian pulses of from $n = 6\;\text{to}\;10$ with different center frequencies is shown in Fig. \ref{fig:D_PLT}.
	\vspace{-4mm}
	\section{Channel Characteristics}
	\vspace{-0.5mm}
	In this paper, we consider mmWave frequency band from 100 GHz - 325 GHz. This section reviews the channel response and ambient noise affecting the propagation of electromagnetic waves in the mmWave channel and these effects are modeled using radiative transfer theory\cite{FEMSEC}. 
	\vspace{-4mm}
	\subsection{Channel Response}
\begin{figure}
		\vspace{-2mm}
		\centering
		\includegraphics[width=0.7\columnwidth, height = 2cm]{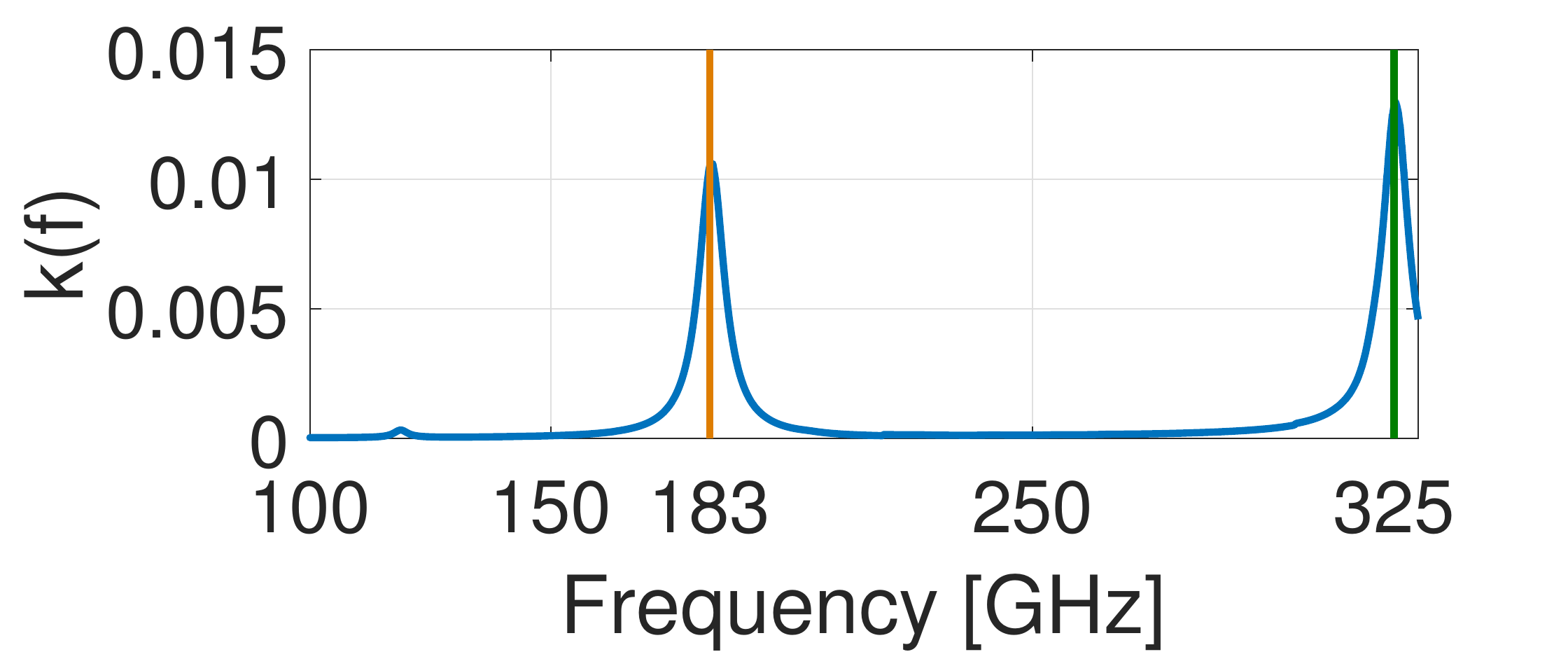}
		\vspace{-3mm}
		\caption{Molecular absorption coefficient for standard summer air mmWave channel.}
		\label{fig:CHANN_COEF}
		\vspace{-7mm}
	\end{figure}

	\vspace{-0.5mm}
	The channel response $H\left( f,d_{r}\right)$ accounts for both spreading loss $H_{spread}\left(f,d_{r} \right) $ and molecular absorption loss $H_{abs}\left(f,d_{r} \right) $ and is represented in frequency domain as 
		\vspace{-1mm}
	\begin{equation}\label{eq:chresp}
	\resizebox{0.5\columnwidth}{!}{$H\left( f,d_{r}\right) = H_{spread}\left( f,d_{r}\right) H_{abs}\left( f,d_{r}\right) $} 
	\vspace{-0.5mm}
	\end{equation}
	\begin{equation}
	\resizebox{0.65\columnwidth}{!}{$H_{spread}\left( f,d_{r}\right) =\left(  \frac{c_{o}}{4\pi d_{r} f_{c}}\right) \exp\left( -\frac{j 2\pi f d_{r}}{c_{o}}\right)$} 
	\vspace{-0.5mm}
	\end{equation}
	\begin{equation}
	\resizebox{0.5\columnwidth}{!}{$H_{abs}\left( f,d_{r}\right) = \exp(-0.5 k\left(f \right)d_{r} )$} 
	\vspace{-0.5mm}
	\end{equation}
	where $f$ denotes frequency, $c_{o}$ is the velocity of light in vacuum, $d_{r}$ is the path length, and $k\left( f\right) $ is the medium absorption coefficient. The medium absorption coefficient $k\left( f\right) $ of the mmWave channel at frequency $f$ composed of $Q$ type molecules is given as
	\vspace{-1mm}
	\begin{equation}
	\vspace{-2mm}
	\resizebox{0.3\columnwidth}{!}{$k\left( f\right)  = \sum\limits_{q=1}^{Q}x_{q}K_{q}\left( f\right).$}
	\vspace{-0.5mm}
	\end{equation} 
	where $x_{q}$ is the mole fraction of molecule type $q$ and $K_{q}$ is the absorption coefficient of individual molecular species. The absorption coefficient $k\left(f \right)$ of the mmWave channel for standard summer air with $1.86 \%$ concentration of water vapor is shown in Fig.\ref{fig:CHANN_COEF}.
	\vspace{-2mm}
	\subsection{Molecular Absorption Noise}
	For transceivers fabricated using graphene material, the effect of thermal noise is negligible\cite{FEMSEC}. The ambient noise affecting the propagation of waves in mmWave band arises due to the propagating of wave itself and is defined as molecular absorption noise. The total molecular absorption noise p.s.d. $S_{N}\left( f,d_{r}\right) $ affecting the transmitted pulse is the sum of background atmospheric noise p.s.d \(S_{N_{B}}\left( f,d_{r}\right)\) and the self induced noise p.s.d. $S_{N_{G}}\left(f,d_{r} \right)$ and is given as\cite{FEMSEC}
	\begin{equation}
	\resizebox{0.27\textwidth}{!}{$S_{N}\left( f,d_{r}\right)  = S_{N_{B}}\left(f,d_{r} \right)+S_{N_{G}}\left(f,d_{r} \right) $} 
	\vspace{-1.5mm}
	\end{equation}
	\begin{equation}\label{eqn:mnm1}
	\vspace{-1mm}
	\resizebox{0.41\textwidth}{!}{$S_{N_{B}}(f, d_{r}) = \lim\limits_{d_{r} \rightarrow \infty} k_{B} T_{0}\left( 1-\exp\left( -k\left(f \right)d_{r} \right) \right) \left( \frac{c_{0}}{\sqrt{4\pi}f_{c}}\right)^{2}$} 
	\vspace{-0.5mm}
	\end{equation}
	\begin{equation}\label{eqn:mnm2}
	\vspace{-0.5mm}
	\resizebox{0.40\textwidth}{!}{$S_{N_{G}}\left(f,d_{r} \right) = S_{G}\left( f\right)\left( 1-\exp\left( -k\left(f \right)d_{r} \right) \right) \left( \frac{c_{0}}{4\pi d_{r} f_{c}}\right)^{2} $} 
	\vspace{-0.5mm}
	\end{equation}
	where $k_{B}$ is the Boltzmann constant, $T_{0}$ is the room temperature and $S_{G}\left( f\right) $ represents p.s.d. of transmitted pulse.  
	\vspace{-6mm}
	\section{System Model}
\begin{figure}[t]
		\centering
		\includegraphics[width=0.8\columnwidth, height = 1.25cm]{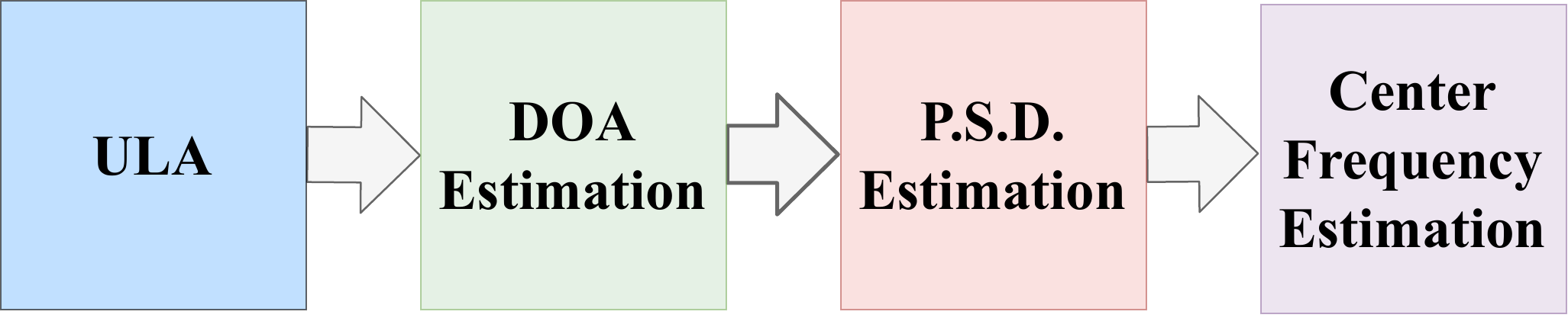}
		\vspace{-2mm}
		\caption{Proposed Center Frequency Estimator.}
		\vspace{-4.5mm}
		\label{fig:BLK_DIAG}
		\vspace{-1mm}
	\end{figure}

	The proposed system model for mmWave device localization and event classification using higher order Gaussian pulses is shown in Fig. \ref{fig:BLK_DIAG}. The mmWave device is localized by estimation of its DOA. The estimation of DOA using the MUSIC algorithm for Gaussian pulses with different center frequencies within terahertz band is investigated in \cite{NANO}. Later, the center frequency classification based on spectral centroid was addressed in \cite{GLOBE}. For a specific order of Gaussian pulse, different events (or status information) at the mmWave device are encoded with different center frequencies that have nonoverlapping half power bandwidth within 100 GHz - 325 GHz. The receiver in Fig. \ref{fig:BLK_DIAG} has the prior knowledge about the different center frequencies transmitted by the mmWave device. The steps for mmWave device localization and event classification is described as follows\\
	$\bullet$ Localize the mmWave node by estimating its DOA.\\
	$\bullet$ Using the estimated DOA, estimate the p.s.d. of the transmitted higher order Gaussian pulse.\\
	$\bullet$ The center frequency of higher order Gaussian pulse is estimated by computing spectral centroid from the estimated p.s.d..\\
	$\bullet$ Based on the estimated center frequency, a particular event is identified (classified).
	\vspace{-4mm}
	\subsection{mmWave Device Localization}
	\vspace{-0.5mm}
	This section describes the procedure for estimating the DOA of a single mmWave device using a ULA with $N$ antennas elements. The spacing between antenna elements in ULA is $d_{s}$ m. The path length between ULA and mmWave device is represented as $d_{r}$. The received wideband higher order Gaussian pulse at the output of the $i$th element in ULA is represented as \cite{FEMSEC}.  
    \vspace{-1mm}
	\begin{equation}\label{eq:datam}
	y_{i}\left(t, d_{r}\right) =  g_{n}\left(t - \tau_{i} \right) * h\left(t,d_{r} \right) + v_{i}\left( t,d_{r}\right)   
	\vspace{-1mm}
	\end{equation}
	\begin{equation}\label{eq:delay_eq}
	\tau_{i} = \left(i -1\right)d_{s}\sin\left( \theta\right) /c 
	\vspace{-1mm}
	\end{equation}
	where $h\left(t,d_{r}\right) $ is the terahertz channel impulse response between ULA and mmWave  device in $\theta$ direction. $v_{i}$ represents molecular absorption noise created between element $i$ of ULA and mmWave device. $\tau_{i}$ represents the time delay for the pulse arriving at $i^{\text{th}}$ element of ULA with respect to a reference element. The time delay $\tau_{i}$ is measured with respect to a reference element of ULA which is located at the origin. When the signal received at the output of the ULA is observed for sufficiently large snapshot observation time $\Delta T$, the Fourier representation of  \eqref{eq:datam} is given as \cite{OPVT}
	\begin{equation}\label{eq:datafm}
	\vspace{-1mm}
	\begin{split}
	Y_{i}\left( f_{b},d_{r}\right) =  e^{-j2\pi f_{b} \tau_{i}} G_{n}\left( f_{b}\right) H\left( f_{b},d_{r}\right) + V_{i}\left( f_{b},d_{r}\right) ,\\ \: \text{for}\;\; b = 0,\cdots, L
	\end{split}
	\vspace{-1mm}
	\end{equation}
	where $f_{b} $ is the frequency bin and $V_{i}\left( f_{b},d_{r}\right) $ is the Fourier coefficient of molecular absorption noise. Further the output of array is observed for $K$ non-overlapping time interval $\Delta T$ and Fourier coefficients is computed for each time interval. Here, $K$ is called as frequency snapshot number. For mmWave device localization using single pulse, the value of $K$ is set as 1, as the output of ULA is observed for a single $\Delta T$. The number of frequency bins $L$ is given as \cite{OPVT}
\vspace{-2mm}
	\begin{equation}\label{eq:timbw}
	L = \lfloor B \cdot \Delta T \rfloor +1
	\vspace{-1mm}
	\end{equation}
	where $\lfloor \cdot  \rfloor$ is the floor operator, $B$ is the terahertz channel bandwidth and the observation time interval $\Delta T$ is significantly greater than the propagation time across ULA. 
	Now, the Fourier coefficients at frequency $f_{b}$ across $N$ sensors for a $K$ number of frequency snapshots is represented in matrix form as 
	\vspace{-1.5mm}
	\begin{equation}\label{eq:datafm_vec}
	\resizebox{0.35\textwidth}{!}{$\boldsymbol{Y}\left(f_{b},d_{r} \right) =  H\left( f_{b},d_{r}\right) \boldsymbol{a}\left(f_{b},\theta \right) \boldsymbol{G}_{n}\left(f_{b} \right) + \boldsymbol{V}\left( f_{b},d_{r}\right)$} 
	\end{equation}
	where $\boldsymbol{Y}\left(f_{b},d_{r} \right) \in \mathbb{C}^{N \times K}$, $\boldsymbol{V}\left( f_{b},d_{r}\right)\in \mathbb{C}^{N \times K}$ and\\ $\boldsymbol{G}_{n}\left(f_{b} \right)\overset{\Delta}{=}\left[ {G}_{n1}\left(f_{b} \right),\cdots,{G}_{nK}\left(f_{b} \right)\right] $.\\$\boldsymbol{a}\left(f_{b},\theta \right) =\left[1, e^{-j2\pi f_{b} \tau_{1}}, \cdots, e^{-j2\pi f_{b} \tau_{N}}\right]^{T} $ is the array manifold vector. The covariance matrix $\boldsymbol{R_{Y}}\left(f_{b},d_{r} \right)$ of $\boldsymbol{Y}\left(f_{b},d_{r} \right)$ is given as 
	\vspace{-1mm}
	\begin{equation}\label{eq:covmat}
	\resizebox{0.5\columnwidth}{!}{$\boldsymbol{R_{Y}}\left(f_{b},d_{r} \right)  = \mathbb{E}\left[ \boldsymbol{Y}\left(f_{b},d_{r} \right)\boldsymbol{Y}\left(f_{b},d_{r} \right)^{H}\right]$} 
	\end{equation}
	where $\left(\cdot \right) ^{H}$ denotes conjugate transpose and $\mathbb{E}\left[ \cdot\right] $ represents expectation. For $K = 1$ and since the value of molecular absorption coefficient $k\left( f\right) $ is very less in the mmWave band (see Fig. \ref{fig:CHANN_COEF}), \eqref{eq:covmat} is simplified as 
	\vspace{-1mm}
	\begin{equation}\label{eq:stdeqn}
	\resizebox{0.85\columnwidth}{!}{$\boldsymbol{R_{Y}}\left(f_{b},d_{r} \right)   =	\left|G_{n}\left( f_{b}\right) \right|^{2} \left|H\left( f_{b},d_{r}\right) \right|^{2} \boldsymbol{a}\left( f_{b},\theta\right)\boldsymbol{a}\left( f_{b},\theta\right)^{H} +\sigma^{2}\left( f_{b},d_{r}\right)  \boldsymbol{I}_{N}$}
	\end{equation}
	In \eqref{eq:stdeqn},  $\boldsymbol{I}_{N}$ is the identity matrix of size $N\times N$ and the term \resizebox{0.5\columnwidth}{!}{$\mathbb{E}\left[ \boldsymbol{V}\left( f_{b},d_{r}\right)\boldsymbol{V}^{H}\left( f_{b},d_{r}\right) \right] =\sigma^{2}\left( f_{b},d_{r}\right)$} is the noise variance around narrow frequency sub-band centered at frequency $f_{b}$. Eqn. \eqref{eq:stdeqn} is same as the covariance matrix at the output of ULA assuming noise to be independent of Gaussian pulses emitted by mmWave  devices. $\sigma^{2}\left( f_{b},d_{r}\right)$ is computed as 
	\begin{equation}
	\resizebox{0.45\columnwidth}{!}{$\sigma^{2}\left( f_{b},d_{r}\right) = \int S_{N}(f_{b},d_{r}) df$}
	\end{equation}
	In this paper, incoherent multiple signal classification (IMUSIC) DOA estimation method is used for localizing mmWave devices. The IMUSIC wideband DOA estimation technique is given as \cite{IMUSIC}
	\begin{equation}\label{eq:IMUS}
	\resizebox{0.4\textwidth}{!}{$P_{\text{IMUSIC}}( \hat{\theta}, d_{r})  = \sum\limits_{b=1}\limits^{L}\frac{\boldsymbol{a}^{H}\left(f_{b},\theta \right)\boldsymbol{a}\left(f_{b},\theta \right) }{\boldsymbol{a}^{H}\left(f_{b},\theta \right)\boldsymbol{E}_{n}\left(f_{b},d_{r} \right)\boldsymbol{E}_{n}^{H}\left(f_{b},d_{r} \right)\boldsymbol{a}\left(f_{b},\theta \right)} $}
	\end{equation}	
	where $\boldsymbol{E}_{n}\left(f_{b},d_{r} \right)$  is the  noise eigenvector matrix which is obtained from eigen value decomposition of $\boldsymbol{R_{Y}}\left(f_{b},d_{r} \right) $.  Eqn. \eqref{eq:IMUS} is called as IMUSIC spectrum and it is observed that, the quality of DOA estimate depends on communication distance between mmWave  device and ULA. The DOA estimate from IMUSIC spectrum is estimated as 
	\vspace{-2.25mm}
	\begin{equation}\label{eq:tht_est}
	\resizebox{0.55\columnwidth}{!}{$\hat{\theta} \left( d_{r}\right) = \argmax_{\boldsymbol{\theta}}\left[ P_{\text{IMUSIC}}( \hat{\theta}, d_{r})\right] $}
		\vspace{-1mm}
	\end{equation}
	Further, the received covariance matrix at each frequency bin $f_{b}$ is estimated as
	\vspace{1mm}
	\begin{equation}
	\resizebox{0.6\columnwidth}{!}{$\boldsymbol{\hat{R}_{Y}}\left(f_{b},d_{r} \right) = \frac{1}{K}\boldsymbol{Y}\left(f_{b},d_{r} \right) \boldsymbol{Y}^{H}\left(f_{b},d_{r} \right)$}
		\vspace{-0.75mm}
	\end{equation}
	\vspace{-8.5mm}
	\subsection{Power spectral density }
	\vspace{-0.5mm}
This section describes the procedure for estimation of the p.s.d. of the received Gaussian pulse. Since the molecular absorption coefficient is low in the mmWave band, the noise term in \eqref{eq:stdeqn} is neglected and the estimated p.s.d. is given as   
   \vspace{-2mm}
	\begin{equation}\label{eqn:psd_est}
	\resizebox{0.65\columnwidth}{!}{$\hat{S}_{G}\left(f_{b} \right) = \left( \boldsymbol{a} \left(f_{b},\hat{\theta} \right) \right)^{\dagger} \hat{\boldsymbol{R}}_{\boldsymbol{Y}}\left(f_{b}.d_{r} \right)\left( \left( \hat{\boldsymbol{a}} \left(f_{b},\hat{\theta} \right) \right)^{H}\right)^{\dagger}$}  
	\vspace{-1mm}
	\end{equation}
	\vspace{-1mm}
	where $\left( \cdot\right) ^{\dagger}$ represents pseudo inverse operator and $\boldsymbol{a} \left(f_{b},\hat{\theta} \right) $ is the array steering vector computed using DOA estimate $\hat{\theta}$.
	\vspace{-8mm}
	\subsection{Center Frequency Estimation}
	\vspace{-0.5mm}
	We now describe the proposed solution for estimating the center frequency of the Gaussian pulse. It is difficult to locate the center frequency of the Gaussian from the estimated p.s.d, due to the presence of molecular resonance peaks and nonuniform channel characteristics. To overcome this problem, the center frequency of the higher order Gaussian pulse is estimated by computing spectral centroid from the estimated p.s.d. of the received Gaussian pulse. The spectral centroid is defined as the center of mass of amplitude or power spectrum. The spectral centroid is used to efficiently discriminate between different cognitive load levels using speech signal and is defined as \cite{SPEECH}
	\vspace{-2mm}
	\begin{equation}
	\resizebox{0.65\columnwidth}{!}{$f_{cen} = \frac{\sum\limits_{b = 1}\limits^{L}f_{b}\cdot\hat{S}_{G}\left( f_{b}\right)\cdot\Delta f}{\sum\limits_{b = 1}\limits^{L}\hat{S}_{G}\left( f_{b}\right)\cdot\Delta f} = \frac{\sum\limits_{b = 1}\limits^{L}f_{b}\cdot\hat{S}_{G}\left( f_{b}\right)}{\sum\limits_{b = 1}\limits^{L}\hat{S}_{G}\left( f_{b}\right)}$}
	\vspace{-1mm}
	\end{equation}
	here $L$ represents number of frequency bins within mmWave band, $\Delta f$ represents width of frequency bin and $\hat{S}_{G}\left( f_{b}\right)\cdot\Delta f$ represents the estimated power spectrum. Based on the computed spectral centroid $f_{cen}$, the center frequency $f_{c{i}}$ of the transmitted higher order Gaussian pulse is estimated according to the following rule
	\begin{equation}\label{eq:freqest}
	\vspace{-0.5mm}
	\hat{f_{c}} = f_{c_{i}}\; \text{if}\;\left | f_{cen}-f_{c_{i}} \right |\leq \left | f_{cen}-f_{c_{j}} \right |\: \forall \: j\neq i
	\vspace{-0.5mm}
	\end{equation} 
	\vspace{-8mm}
\section{Simulation Results}
\vspace{-0.5mm}
In this section, simulation results are presented to analyze DOA and center frequency estimation accuracy using a single Gaussian pulse with its order varying from $n =6^{\text{th}}\;\text{to}\;10^{\text{th}}$ order in higher mmWave frequency band (100 GHz - 325 GHz). The IMUSIC and Spectral centroid estimation algorithms explained in the previous section is implemented using MATLAB R2014a.
	\vspace{-4mm}
	\subsection{Parameters and Performance metrics}
	\vspace{-0.5mm}
	In the simulation, the number of antenna elements in ULA is considered as 8, as a large number of antenna elements in ULA provides good DOA estimation accuracy. The distance between consecutive antenna elements in ULA is half the wavelength $\lambda_{min}$ of frequency 325 GHz, that is $d_{s} = \SI{0.4615}{\milli\meter}$ to avoid spatial aliasing. The transmitting mmWave device is assumed to be located in the far-field region of ULA with DOA $12.5175^{\circ}$. Two different snapshot observation time intervals $\Delta T = 42\;\si{\pico}s$ (L = 11) and  $\Delta T = 180\;\si{\pico}s$ (L = 42) are considered in the simulation. The power of higher order Gaussian pulses is set to 0.1 \si{\micro}W. Further, the Nyquist sampling frequency required at the ULA will be 650 GHz, which is twice the maximum frequency 325 GHz. But, the maximum sampling frequency of the current data converters is 100 GHz \cite{GBPS}. Using sub-Nyquist methods like the finite rate of innovation, the required Nyquist rate can be reduced by a factor of 20 \cite{CHAN}. The high-resolution transmission molecular absorption (HITRAN) database \cite{HTRAN} is used to obtain the molecular absorption coefficient $k\left( f\right) $ of the mmWave channel. The estimation accuracy of the parameters DOA estimate $\hat{\theta}$  and center frequency $f_{c}$ for single sensor node is measured in terms of root mean square error (RMSE) and is defined as
	\vspace{-2mm}
	\begin{equation}
	\vspace{-0.5mm}
	\resizebox{0.5\columnwidth}{0.07\columnwidth}{$RMSE_{m} = \sqrt{\frac{1}{N_{run}}\sum\limits_{i=1}^{N_{run}}\left (\hat{m}\left ( i \right ) -m \right )^{2}}$}
	\vspace{-2.5mm}
	\end{equation}
	where $N_{run}$ is the total number of single pulse transmissions, $\hat{m}\left ( i \right )\in \left\lbrace \hat{\theta}, \hat{f}_{c}\right\rbrace $ is the estimate of parameter in $i^{\text{th}}$ simulation run and its corresponding true value is $m \in \left\lbrace \theta, f_{c}\right\rbrace $. The classification accuracy of center frequency of higher order Gaussian pulse is defined in terms of true positive rate (TPR). TPR is defined as the number of correct classification of center frequency in total number of single pulse transmissions. The value of  $N_{run}$ is selected as 500.
	\vspace{-4mm}
	\subsection{DOA estimation accuracy}
\begin{figure*}[t]
		\subfigure{
			\includegraphics[width=0.425\columnwidth,height=2.19cm]{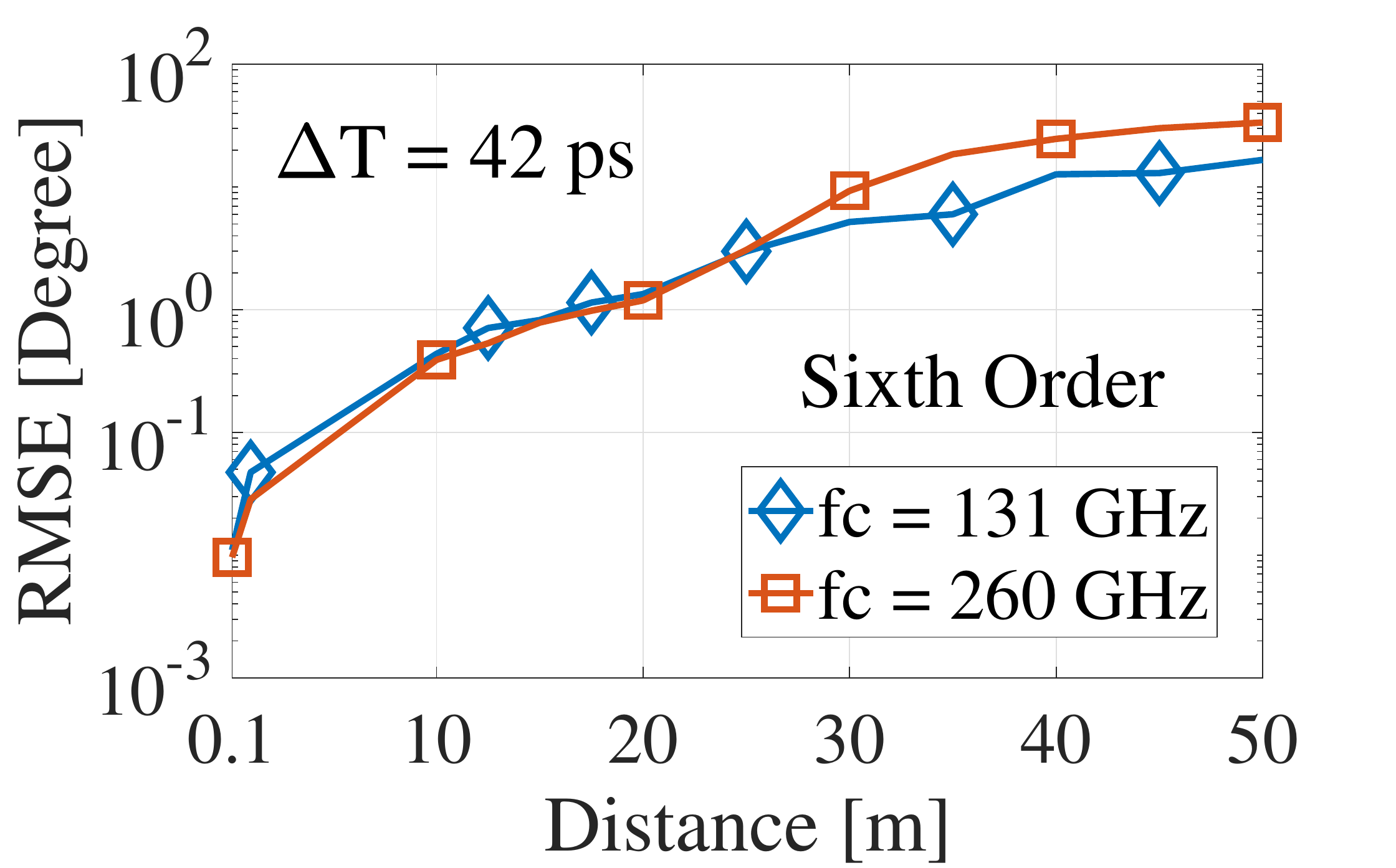}
		}
		\hspace{-5mm}
		\vspace{-0.75mm}
		\subfigure{
			\includegraphics[width=0.425\columnwidth,height=2.19cm]{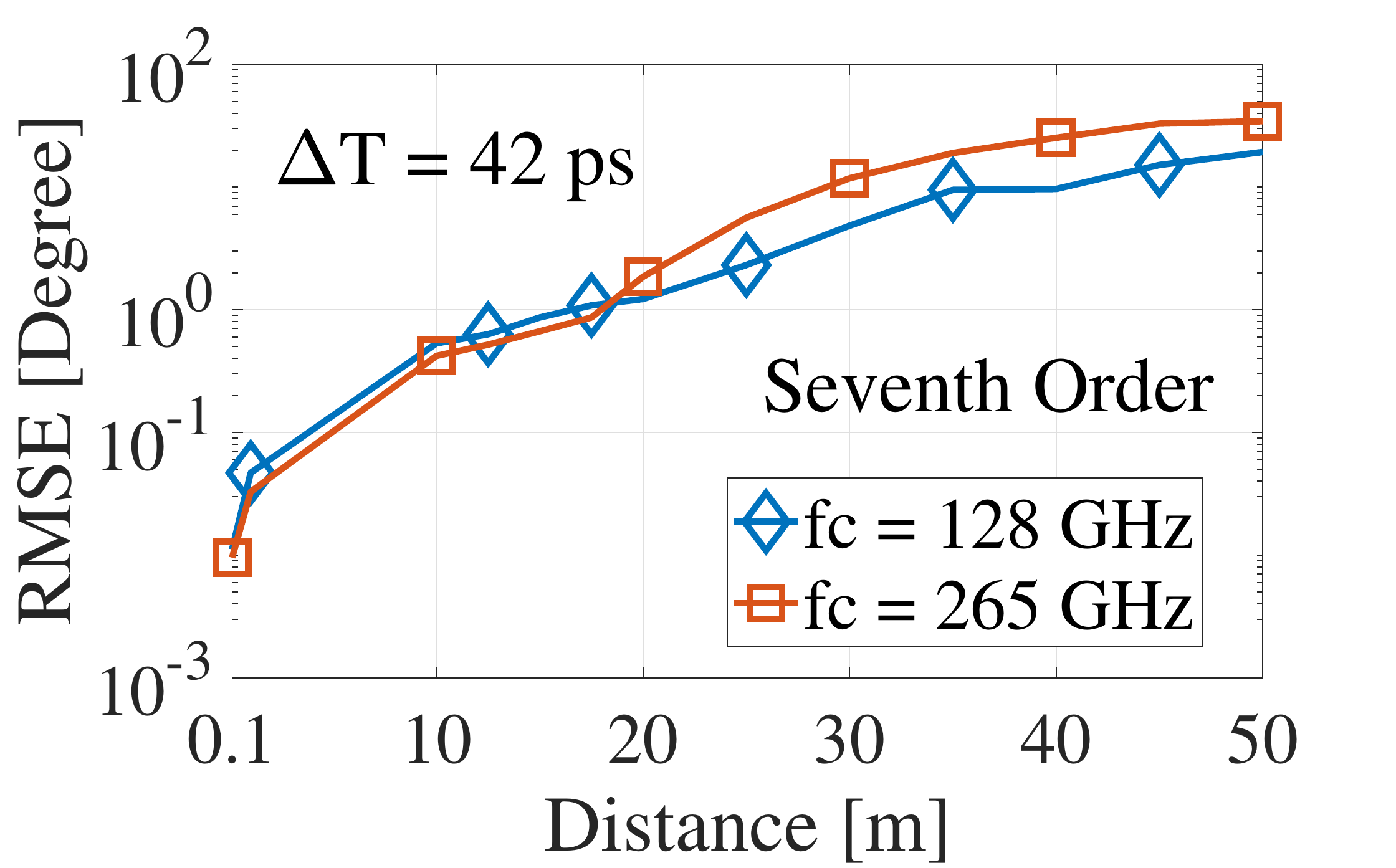}
		}
		\hspace{-5mm}
		\vspace{-0.75mm}
		\subfigure{
			\includegraphics[width=0.425\columnwidth,height=2.19cm]{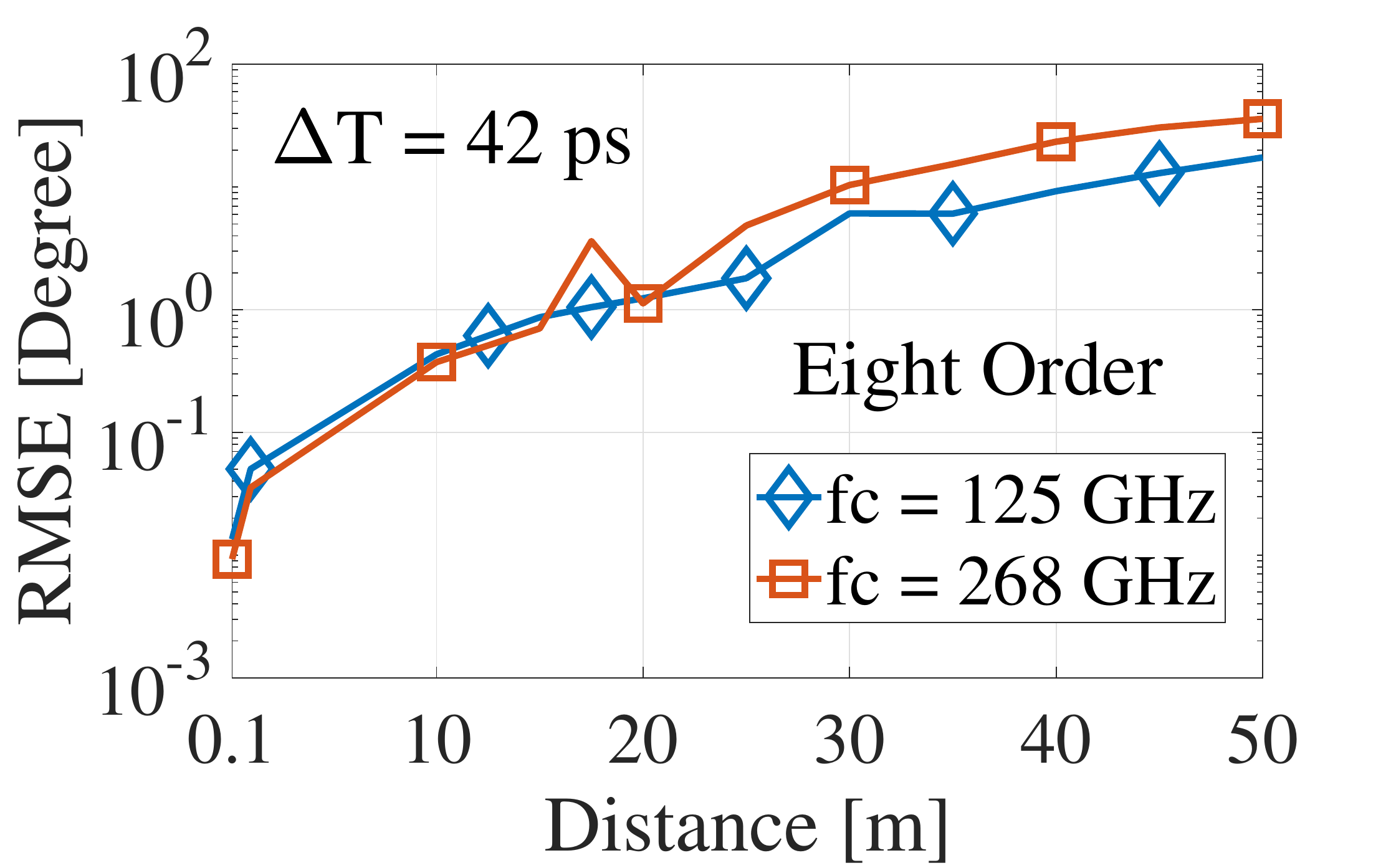}
		}
		\hspace{-5mm}
		\vspace{-0.75mm}
		\subfigure{
			\includegraphics[width=0.425\columnwidth,height=2.19cm]{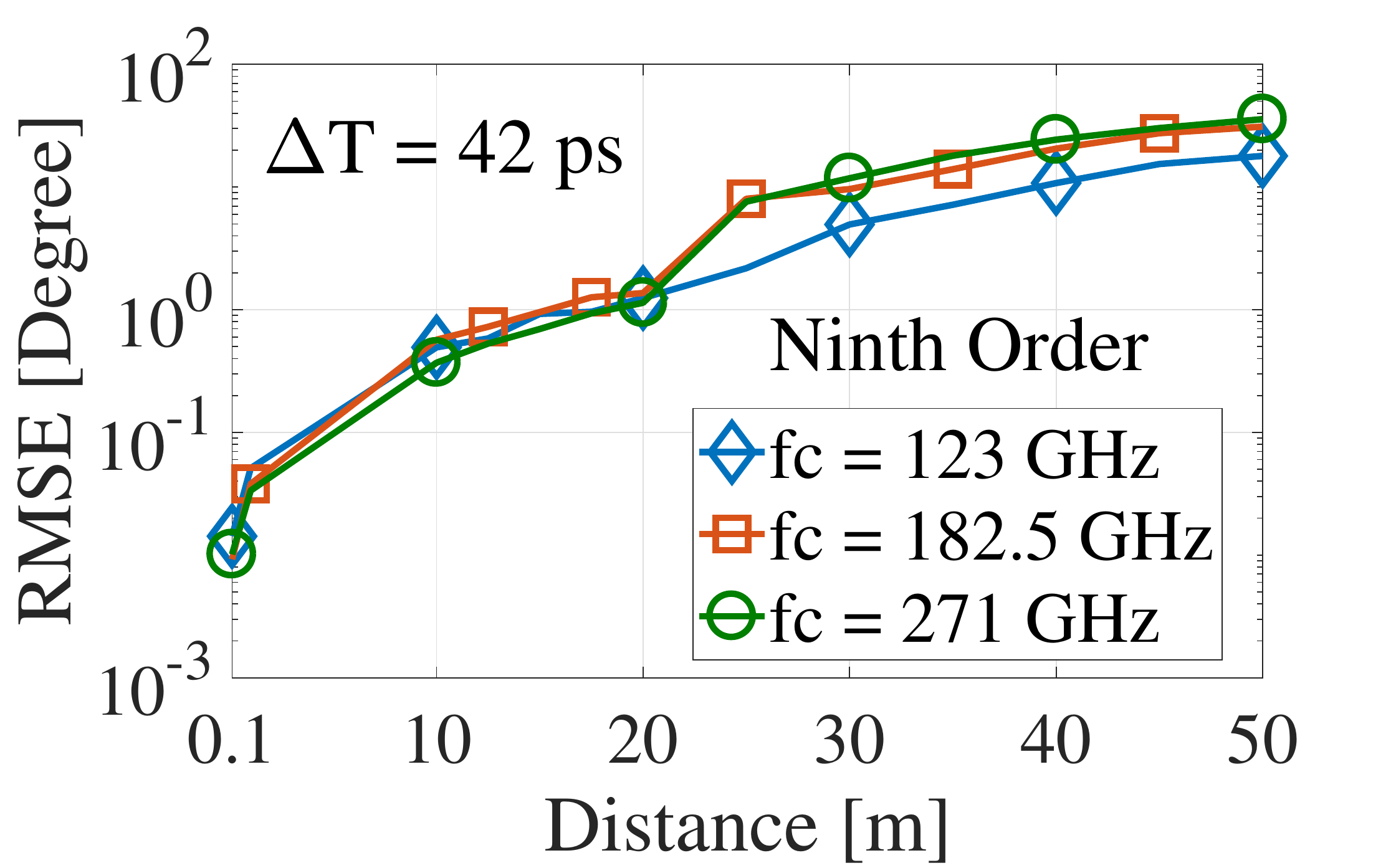}
		}
		\hspace{-5mm}
		\vspace{-0.75mm}
		\subfigure{
			\includegraphics[width=0.425\columnwidth,height=2.19cm]{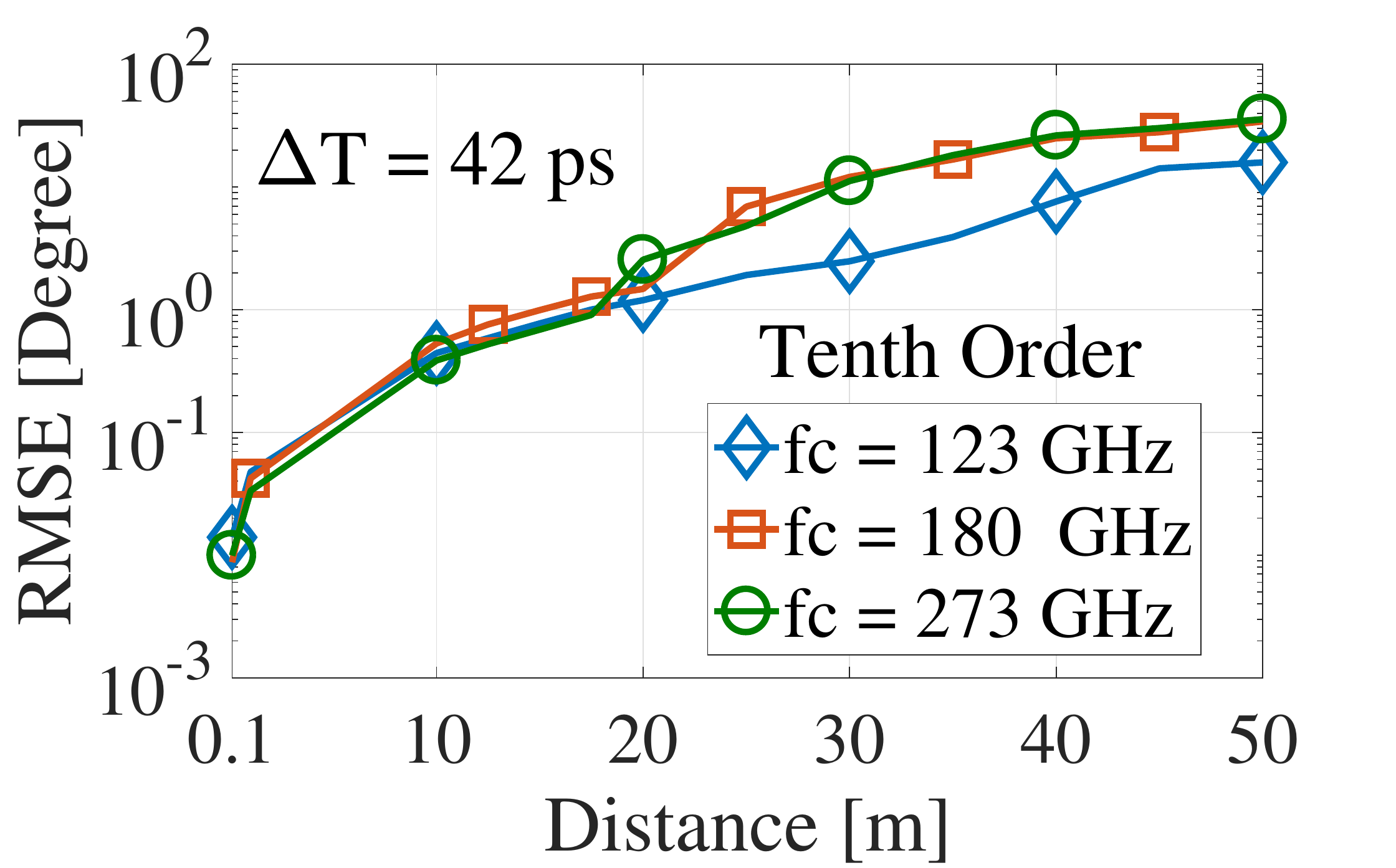}
		}
		\vspace{-0.75mm}
		\\
		\subfigure{
			\includegraphics[width=0.425\columnwidth,height=2.19cm]{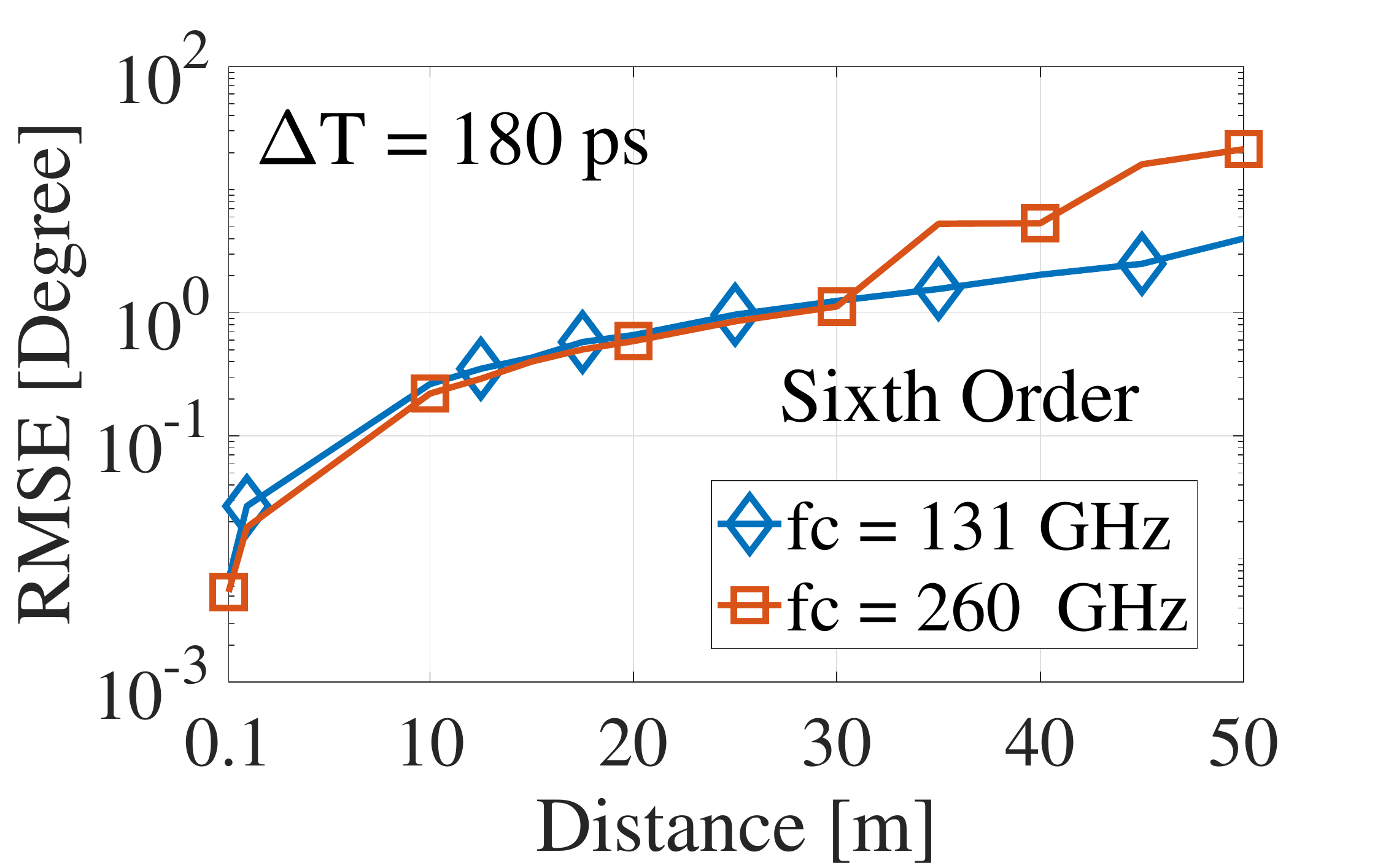}
		}
		\hspace{-5mm}
		\subfigure{
			\includegraphics[width=0.425\columnwidth,height=2.19cm]{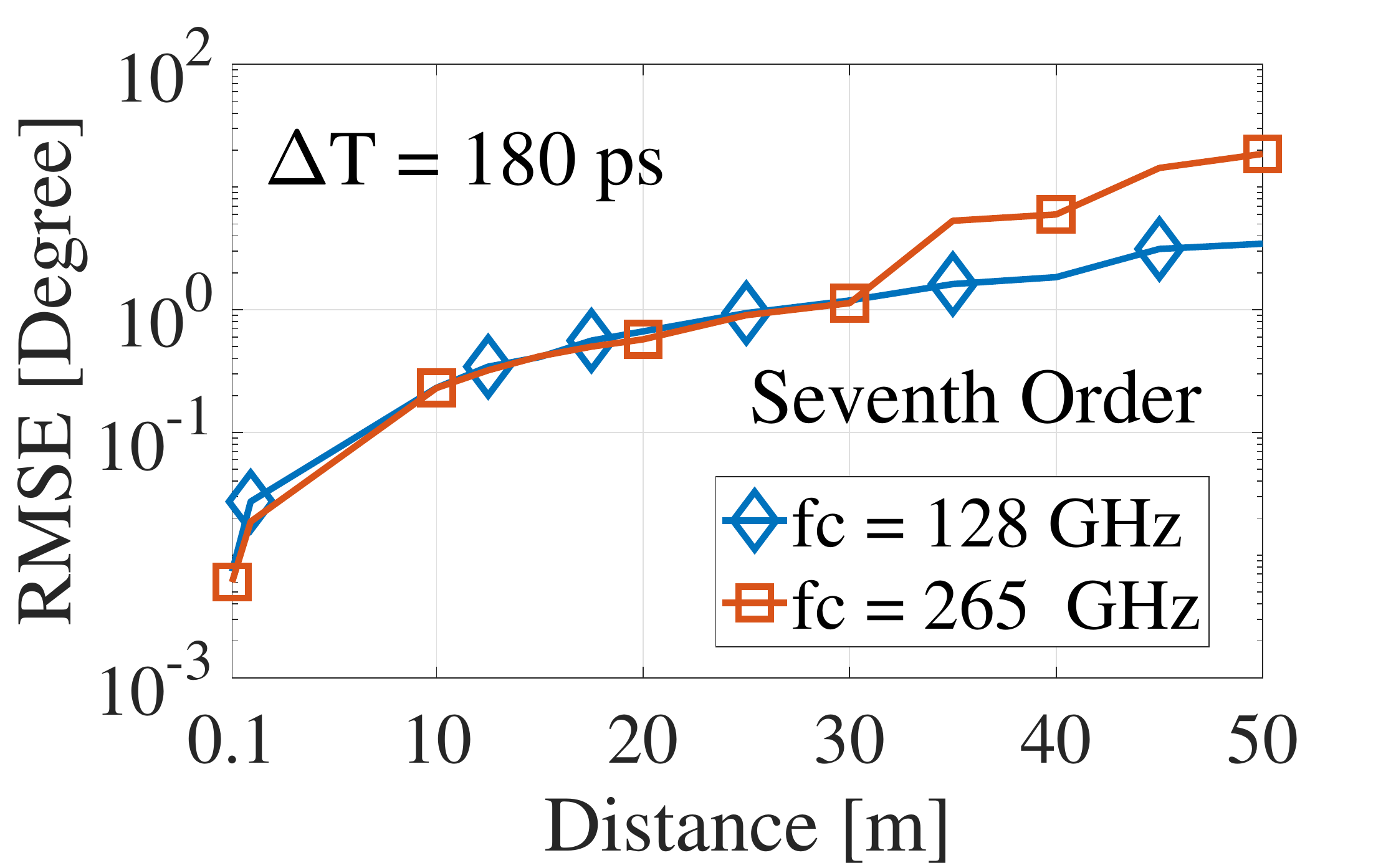}
		}
		\hspace{-5mm}
		\subfigure{
			\includegraphics[width=0.425\columnwidth,height=2.19cm]{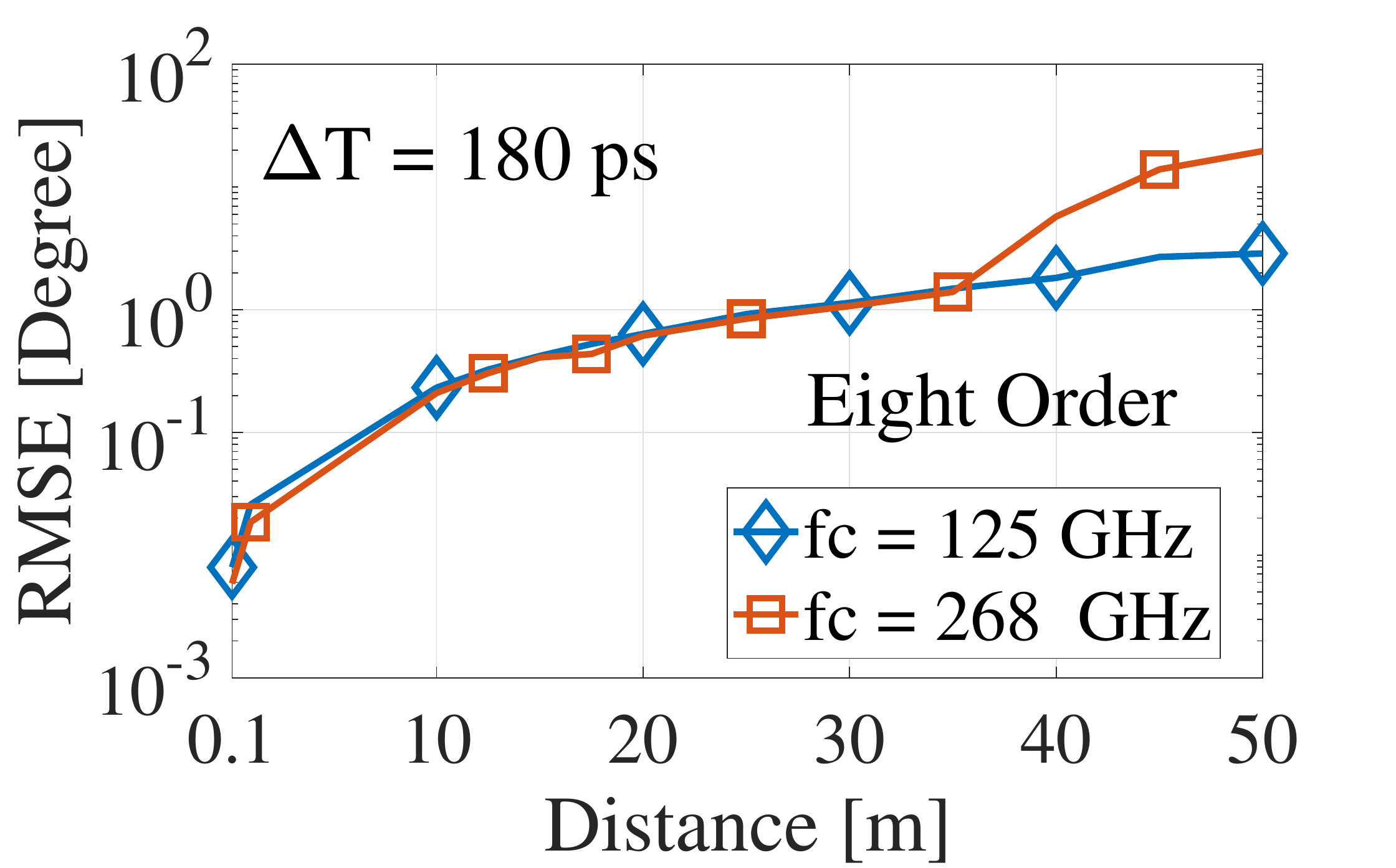}
		}
		\hspace{-5mm}
		\subfigure{
			\includegraphics[width=0.425\columnwidth,height=2.19cm]{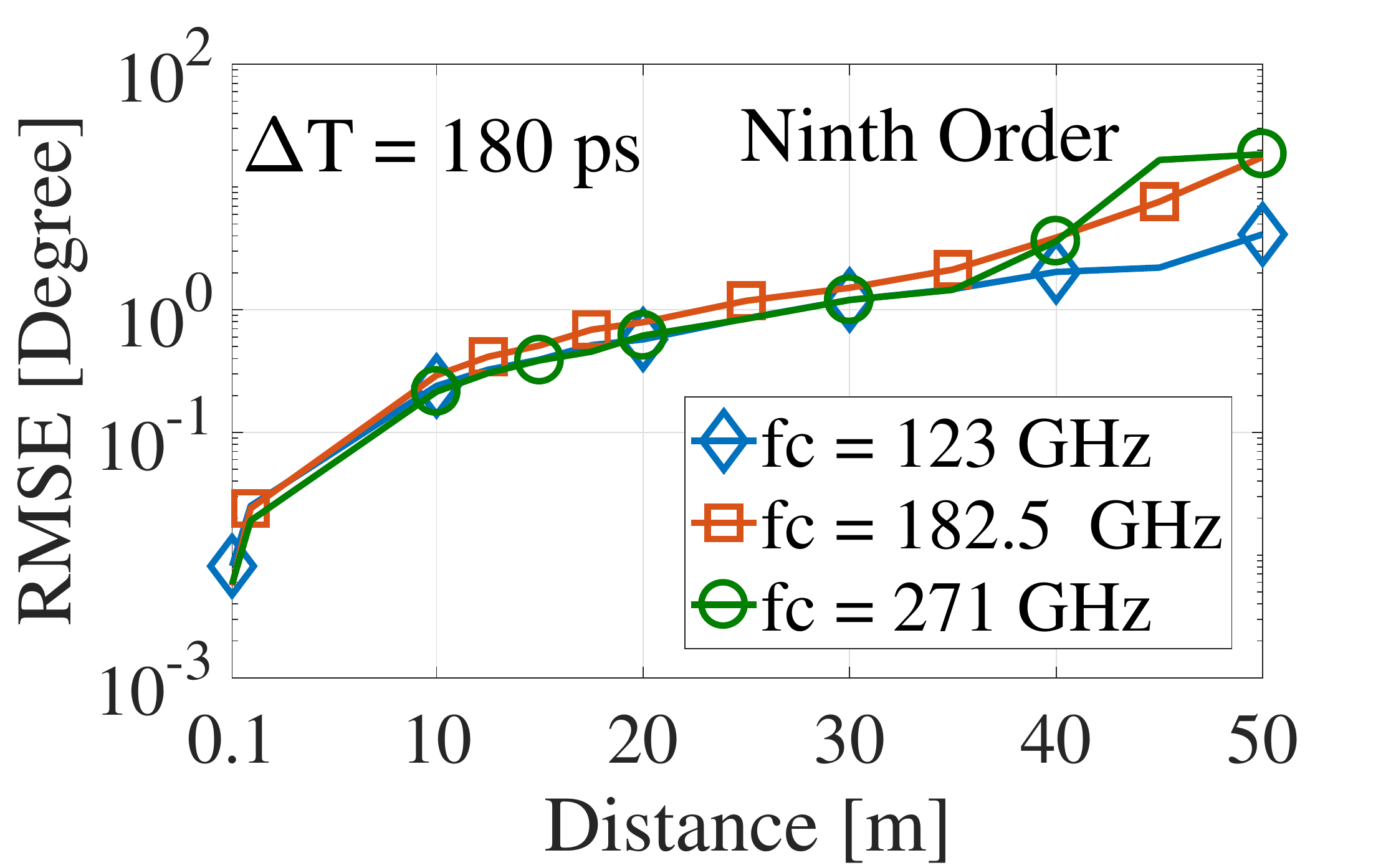}
		}
		\hspace{-5mm}
		\subfigure{
			\includegraphics[width=0.425\columnwidth,height=2.19cm]{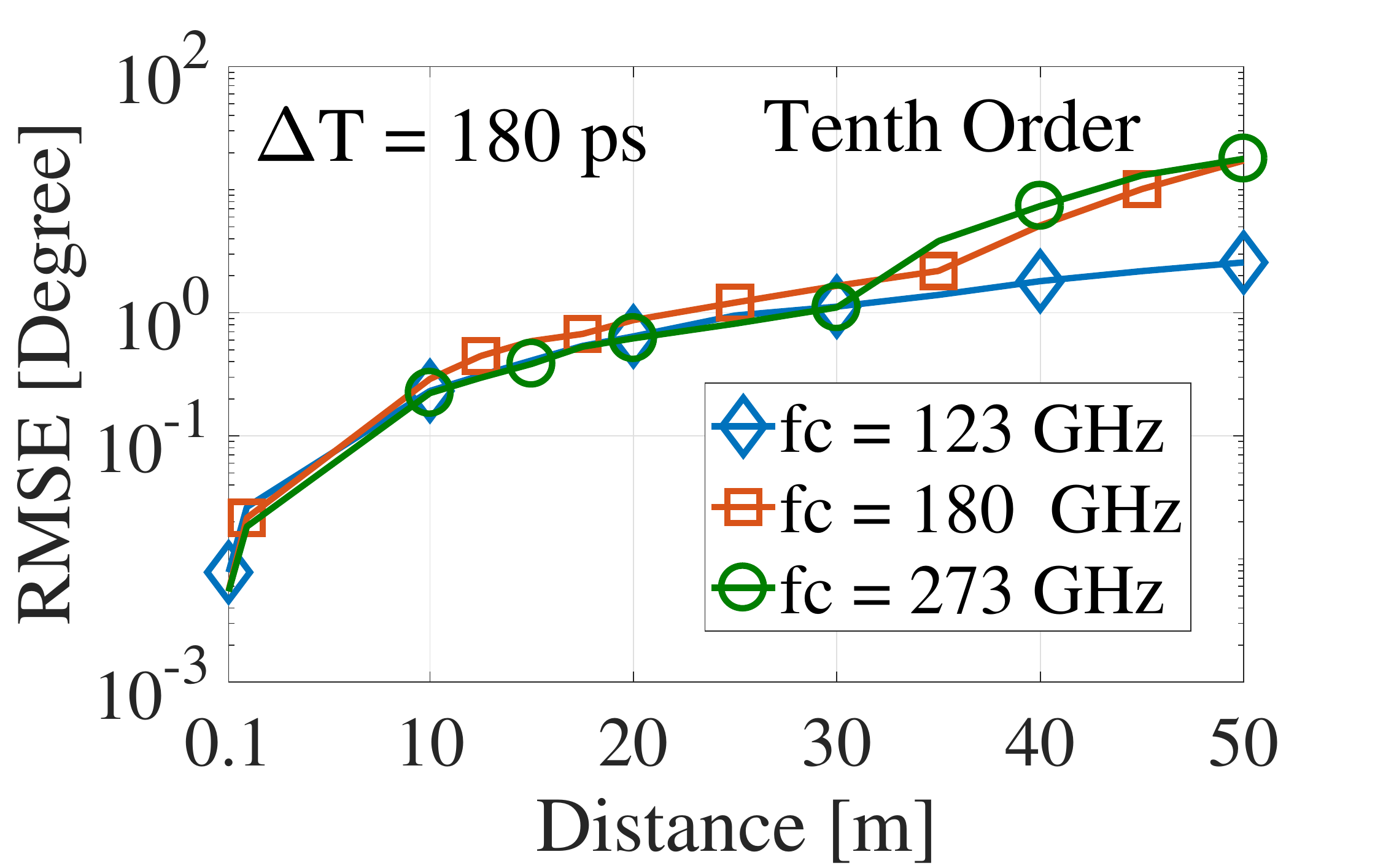}
		}
		\vspace{-5mm}
		\caption{DOA estimation accuracy for Higher order Gaussian pulses }
		\label{fig:DOA_RMSE}
		\vspace{-1mm}
	\end{figure*}

	\vspace{-0.5mm}
	The DOA estimation performance as a function of path length is shown in Fig. \ref{fig:DOA_RMSE} and it is observed that DOA estimation accuracy decreases with an increase in path length. Further, we make the following observations: \textbf{1)} For a given pulse power, lower center frequency Gaussian pulses which have higher energy content provides better DOA estimation accuracy at large path lengths as compared to low-energy higher center frequency Gaussian pulses (see Table \ref{tabl:ERG}). A possible explanation for this outcome is because the signal to noise ratio at each frequency bin does not depend on the pulse duration. \textbf{2)} The DOA estimation performance is better for large snapshot observation time  $\Delta T = 180\; \si{\pico}s$ as the RMSE is below 1 degree for path length up to 30 m as compared to 20 m for $\Delta T = 42 \;\si{\pico}s$. This is due to availability of large number of frequency bins at $\Delta T = 180\; \si{\pico}s$.
	\vspace{-4mm}
	\subsection{Center Frequency Classification}
\begin{figure*}
		\vspace{-5mm}
		\subfigure{
			\includegraphics[width=0.425\columnwidth,height=2.19cm]{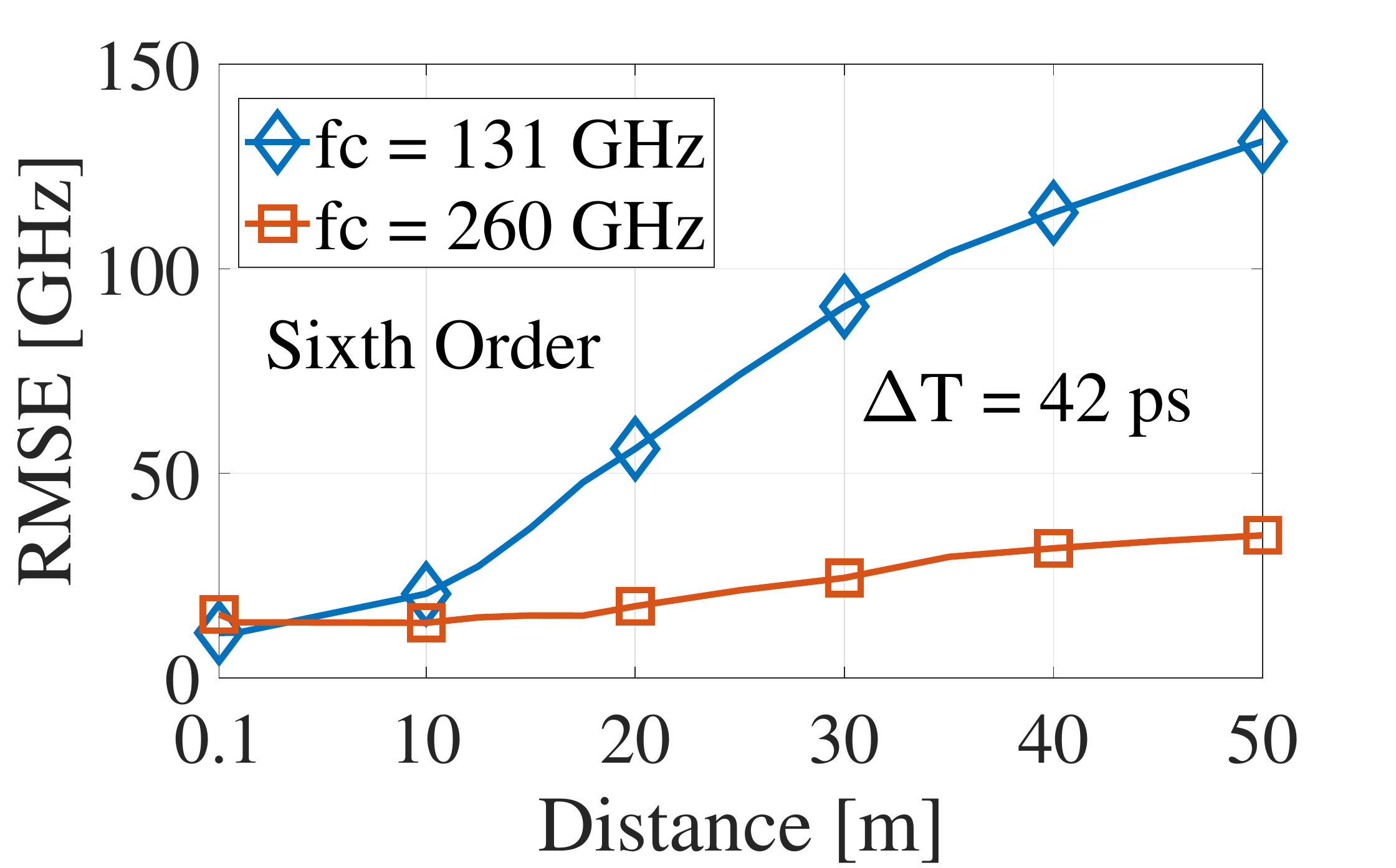}
		}
		\hspace{-5mm}
		\vspace{-0.75mm}
		\subfigure{
			\includegraphics[width=0.425\columnwidth,height=2.19cm]{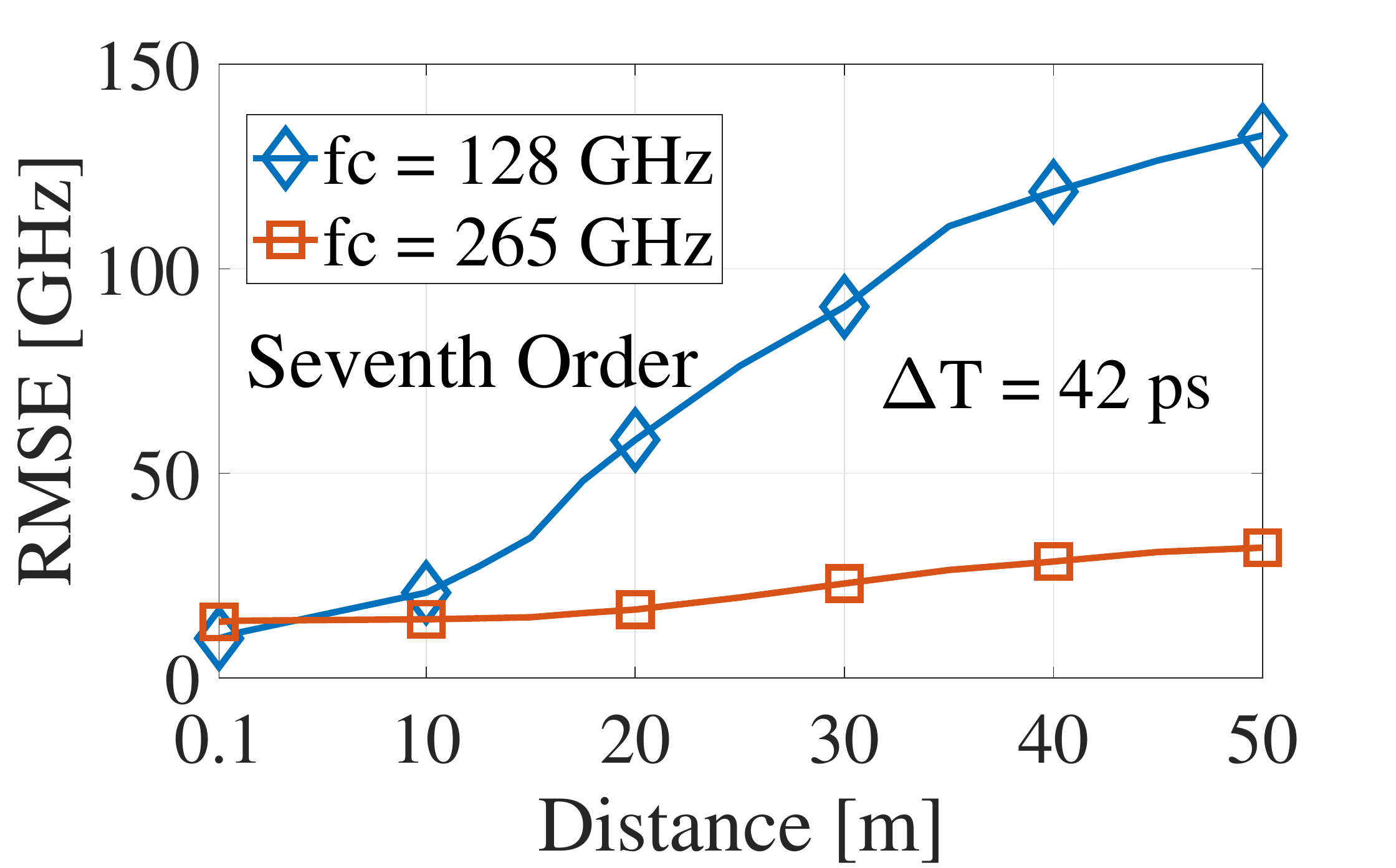}
		}
		\hspace{-5mm}
		\vspace{-0.75mm}
		\subfigure{
			\includegraphics[width=0.425\columnwidth,height=2.19cm]{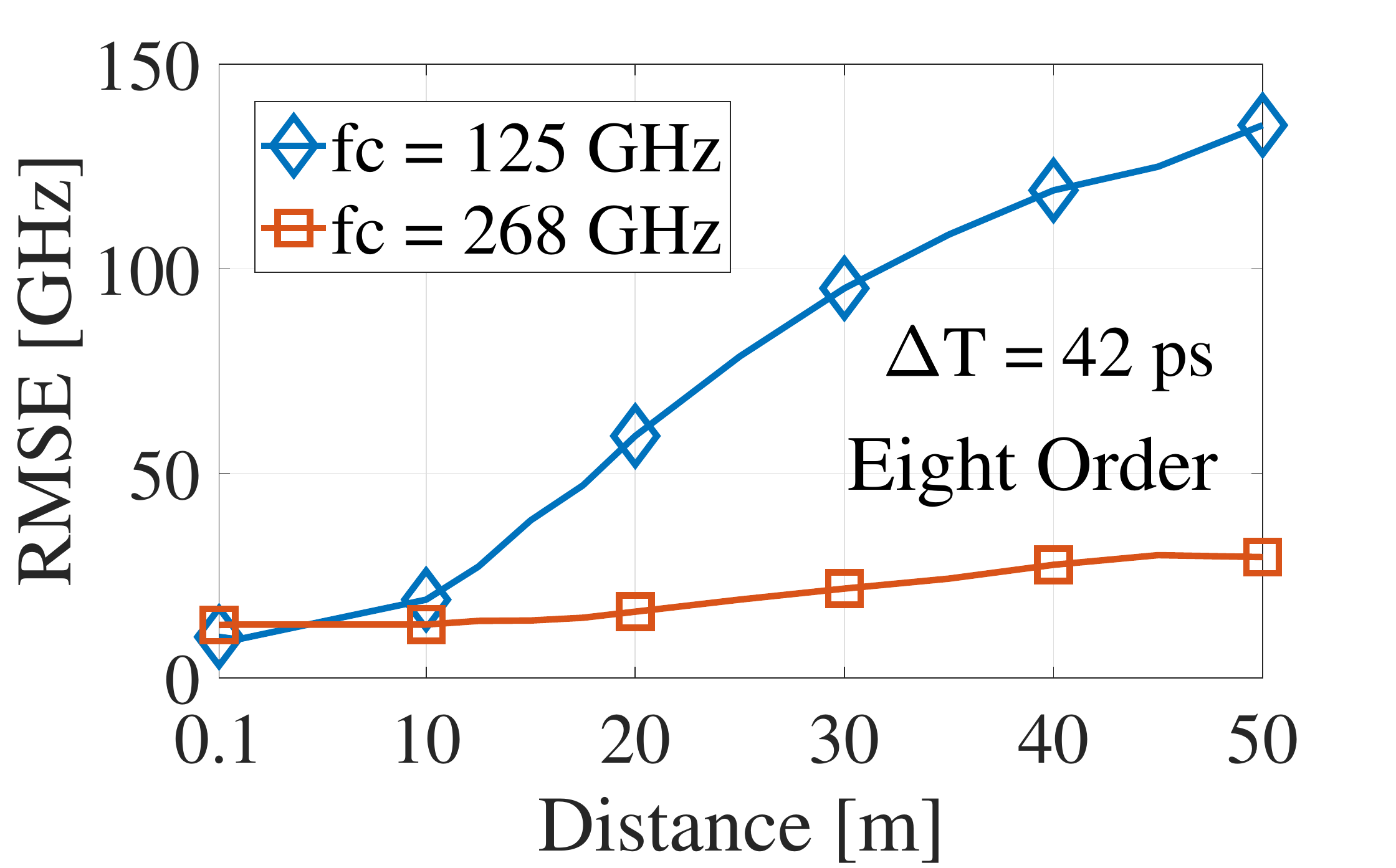}
		}
		\hspace{-5mm}
		\vspace{-0.75mm}
		\subfigure{
			\includegraphics[width=0.425\columnwidth,height=2.19cm]{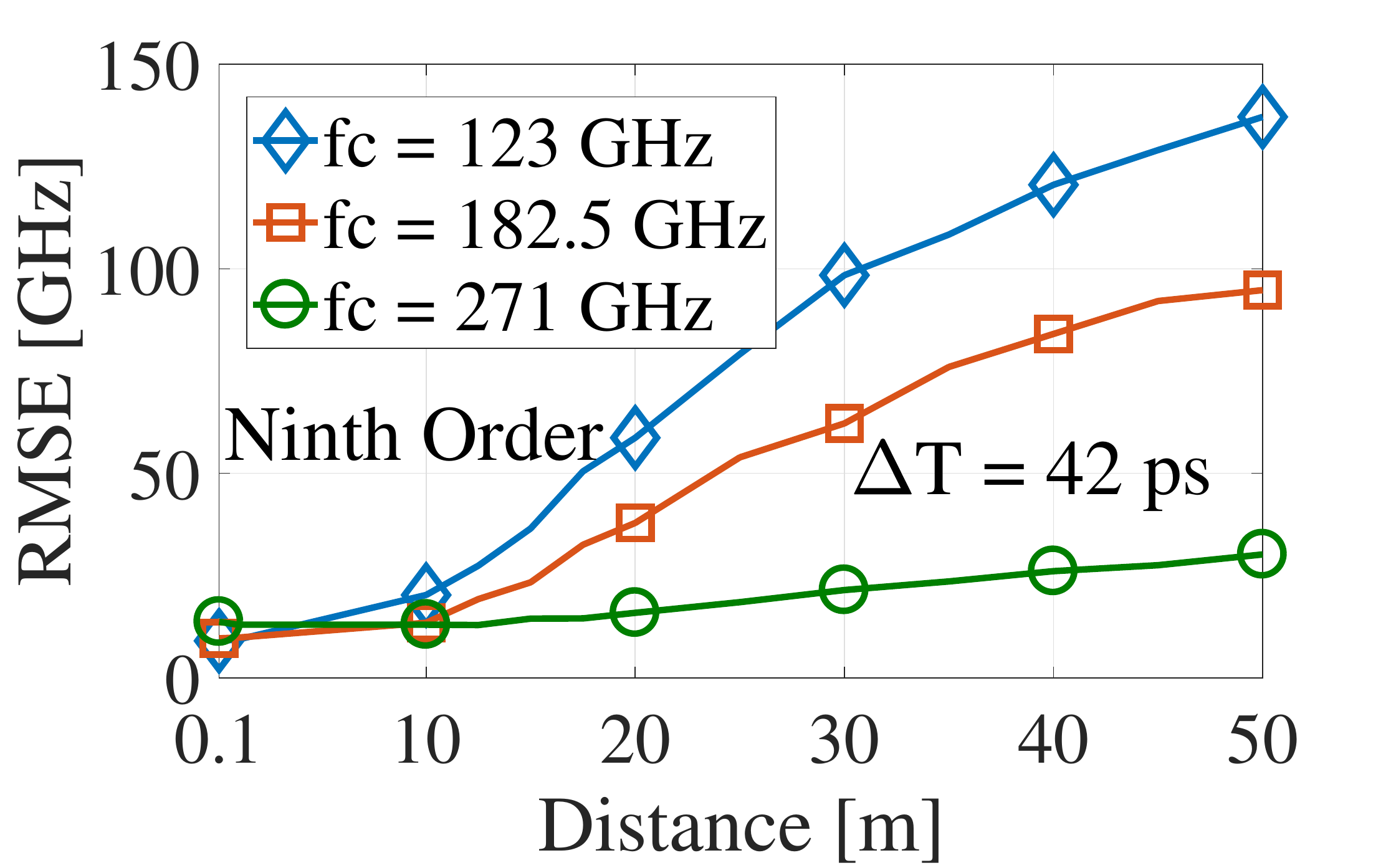}
		}
		\hspace{-5mm}
		\vspace{-0.75mm}
		\subfigure{
			\includegraphics[width=0.425\columnwidth,height=2.19cm]{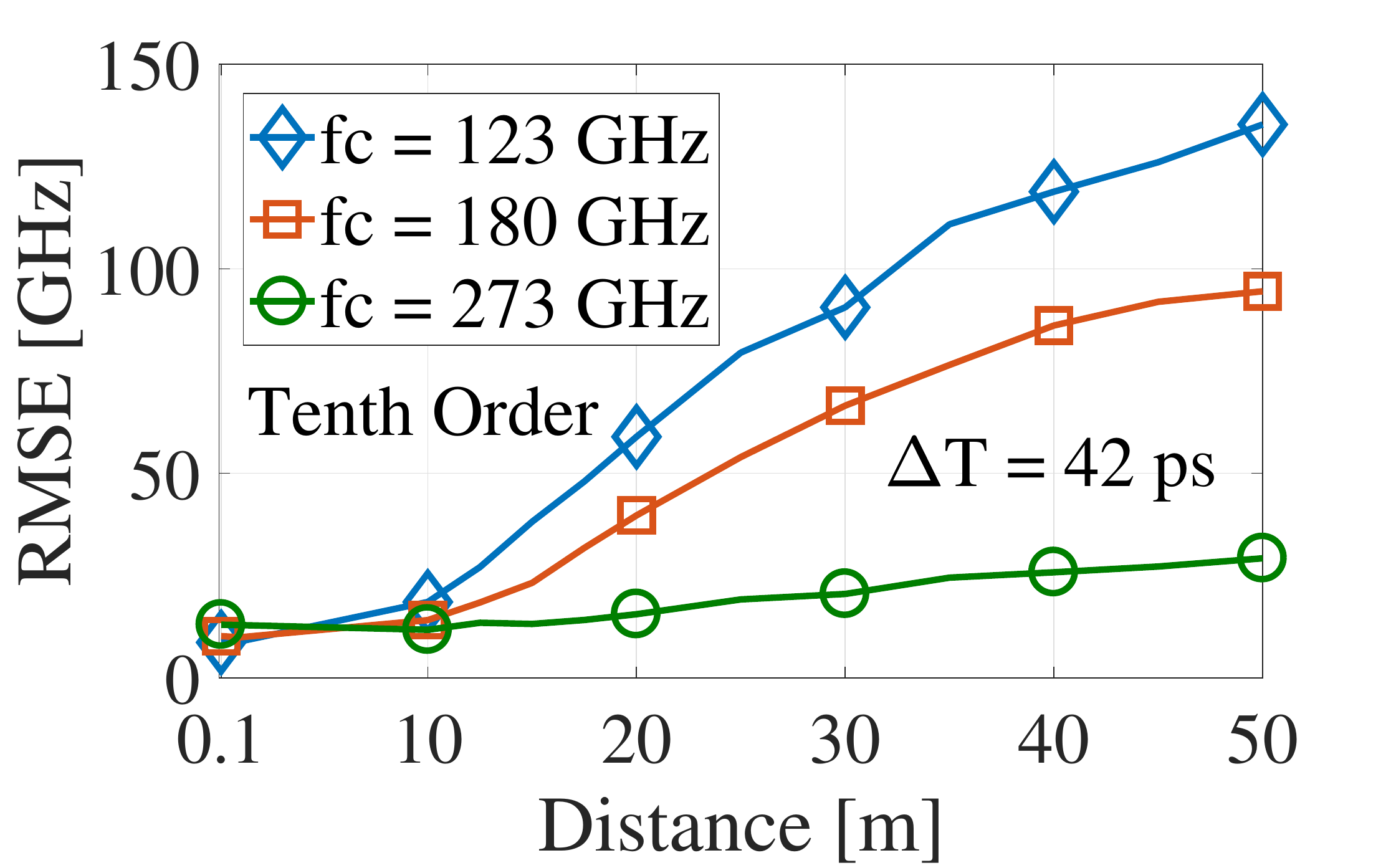}
		}
		\vspace{-0.75mm}
		\\
		\subfigure{
			\includegraphics[width=0.425\columnwidth,height=2.19cm]{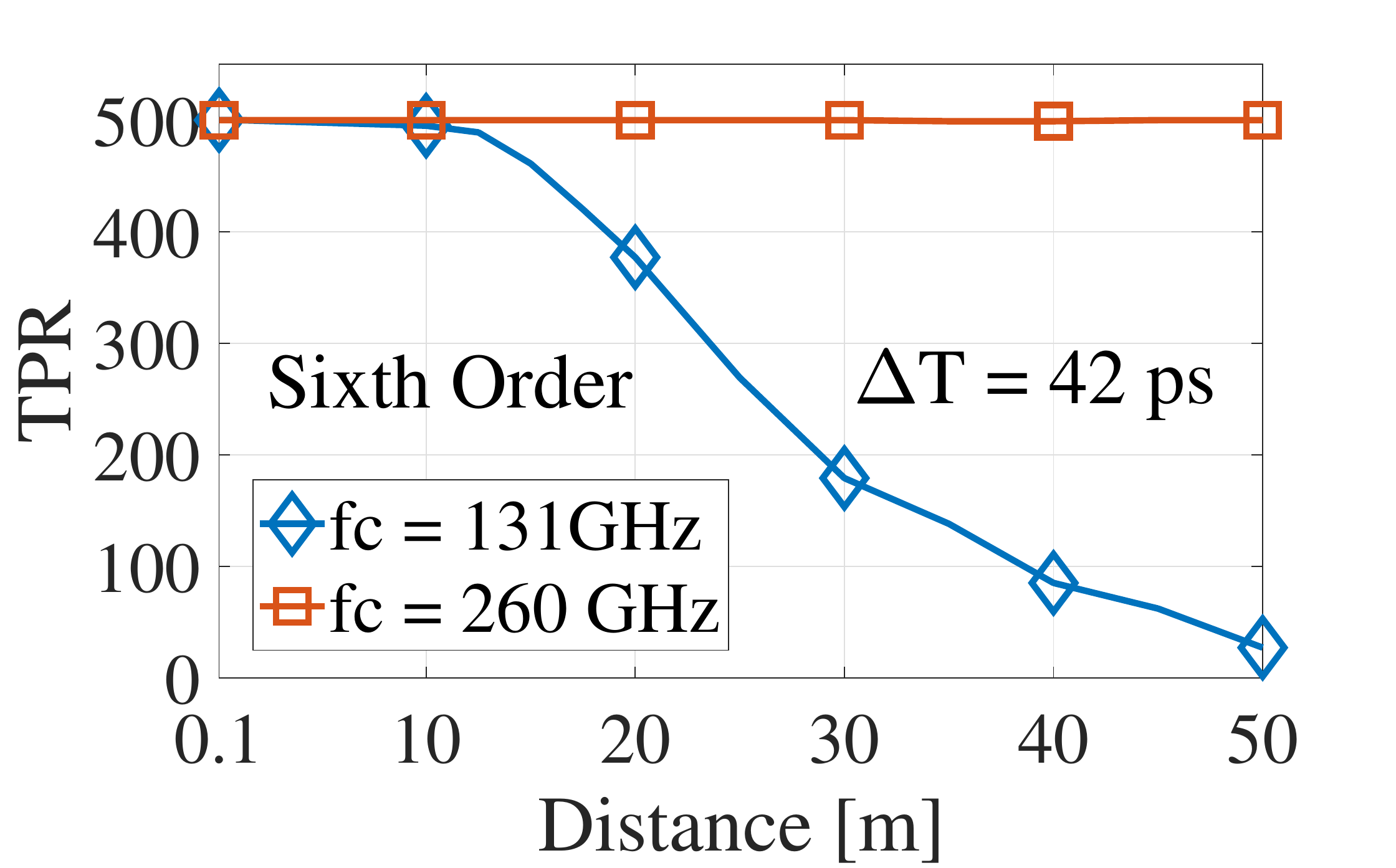}
		}
		\hspace{-5mm}
		\vspace{-0.75mm}
		\subfigure{
			\includegraphics[width=0.425\columnwidth,height=2.19cm]{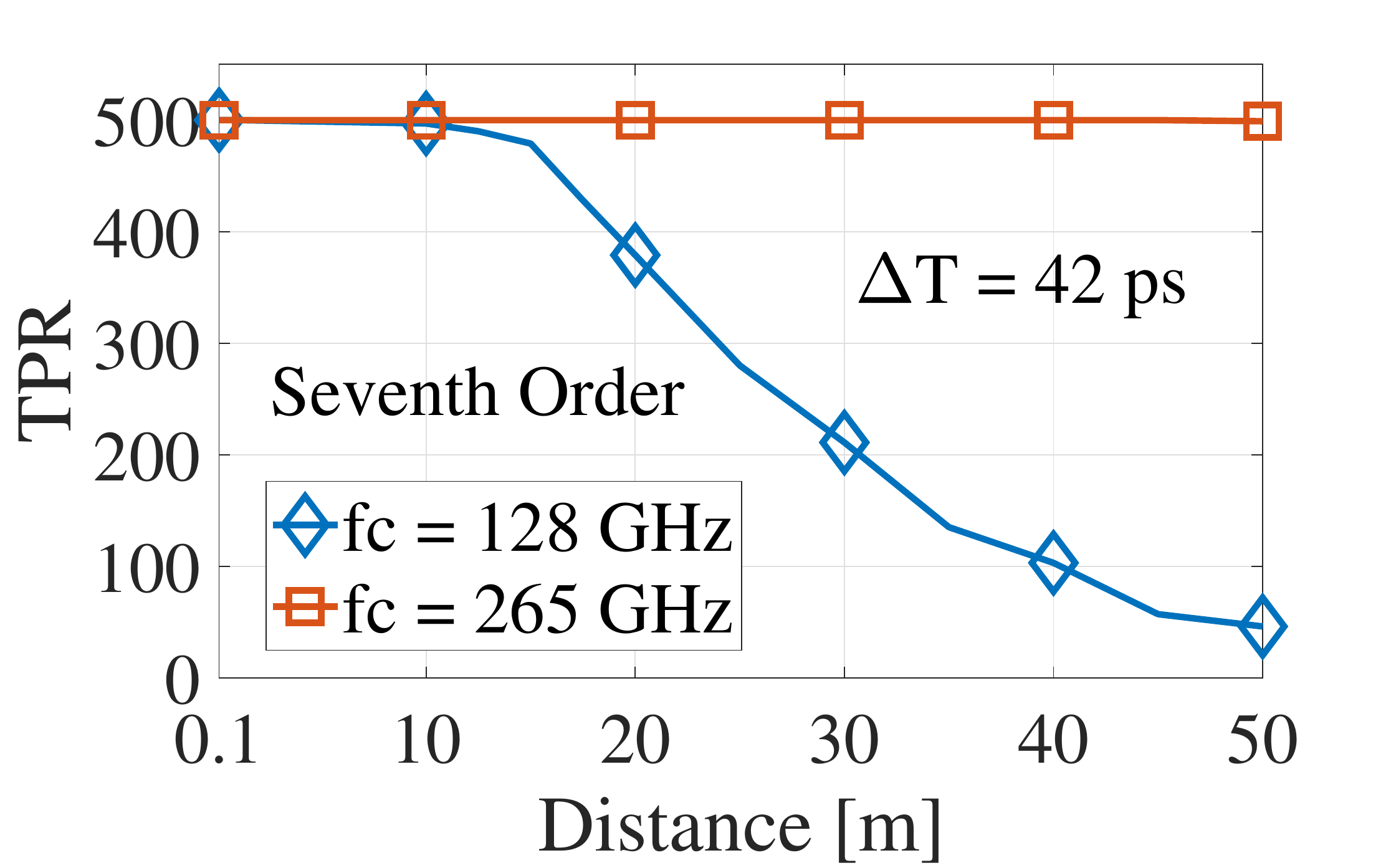}
		}
		\hspace{-5mm}
		\vspace{-0.75mm}
		\subfigure{
			\includegraphics[width=0.425\columnwidth,height=2.19cm]{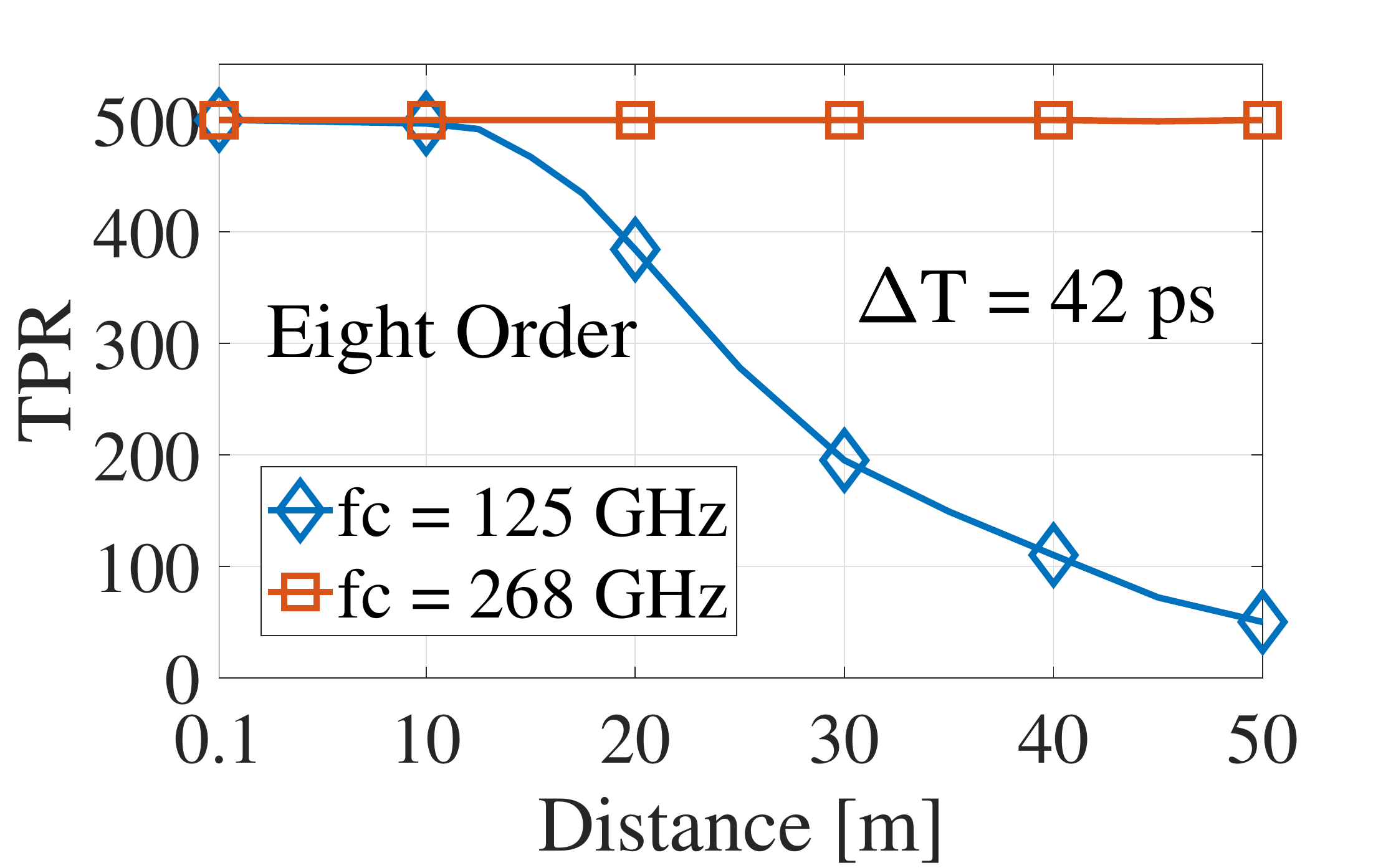}
		}
		\hspace{-5mm}
		\vspace{-0.75mm}
		\subfigure{
			\includegraphics[width=0.425\columnwidth,height=2.19cm]{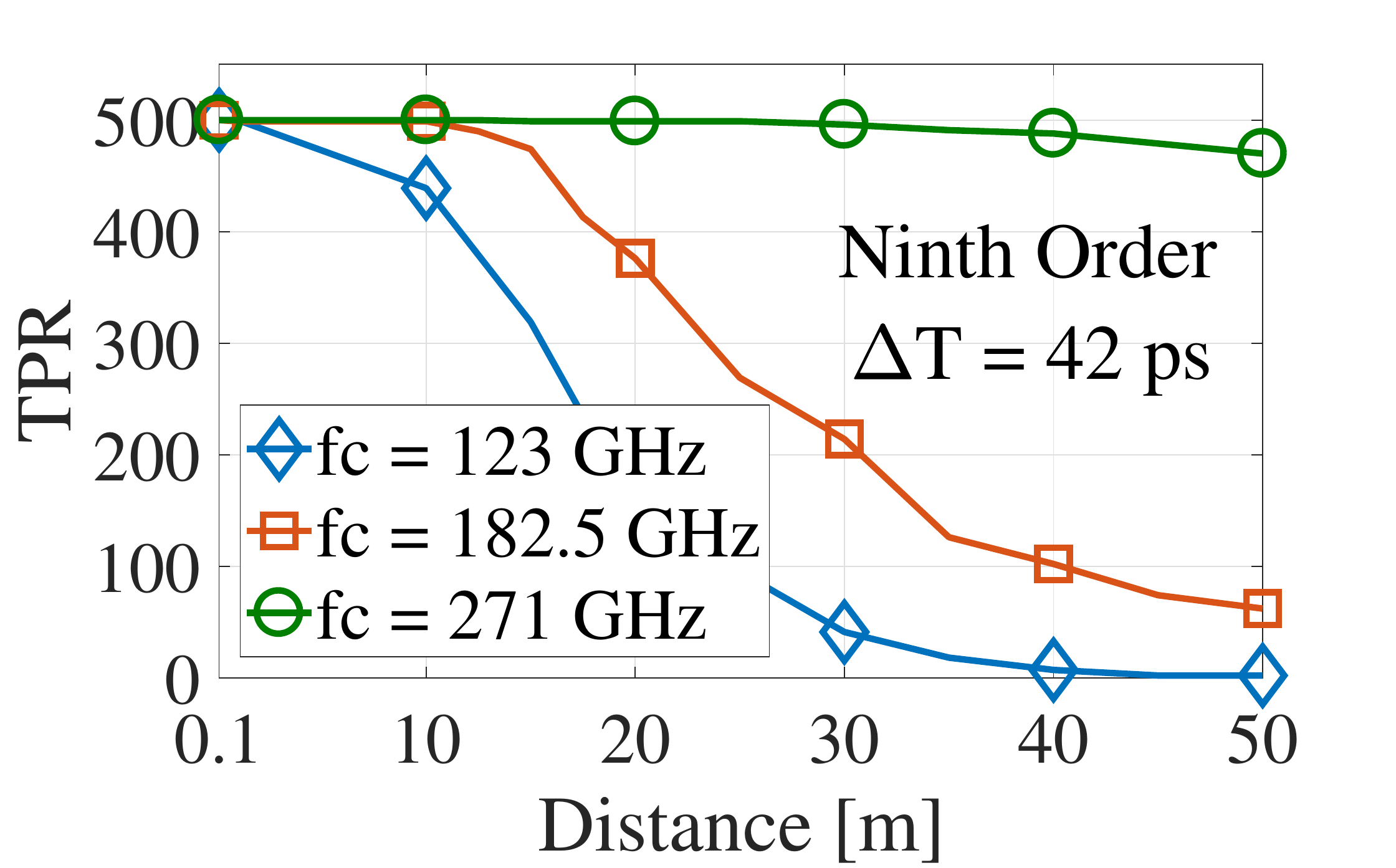}
		}
		\hspace{-5mm}
		\vspace{-0.75mm}
		\subfigure{
			\includegraphics[width=0.425\columnwidth,height=2.19cm]{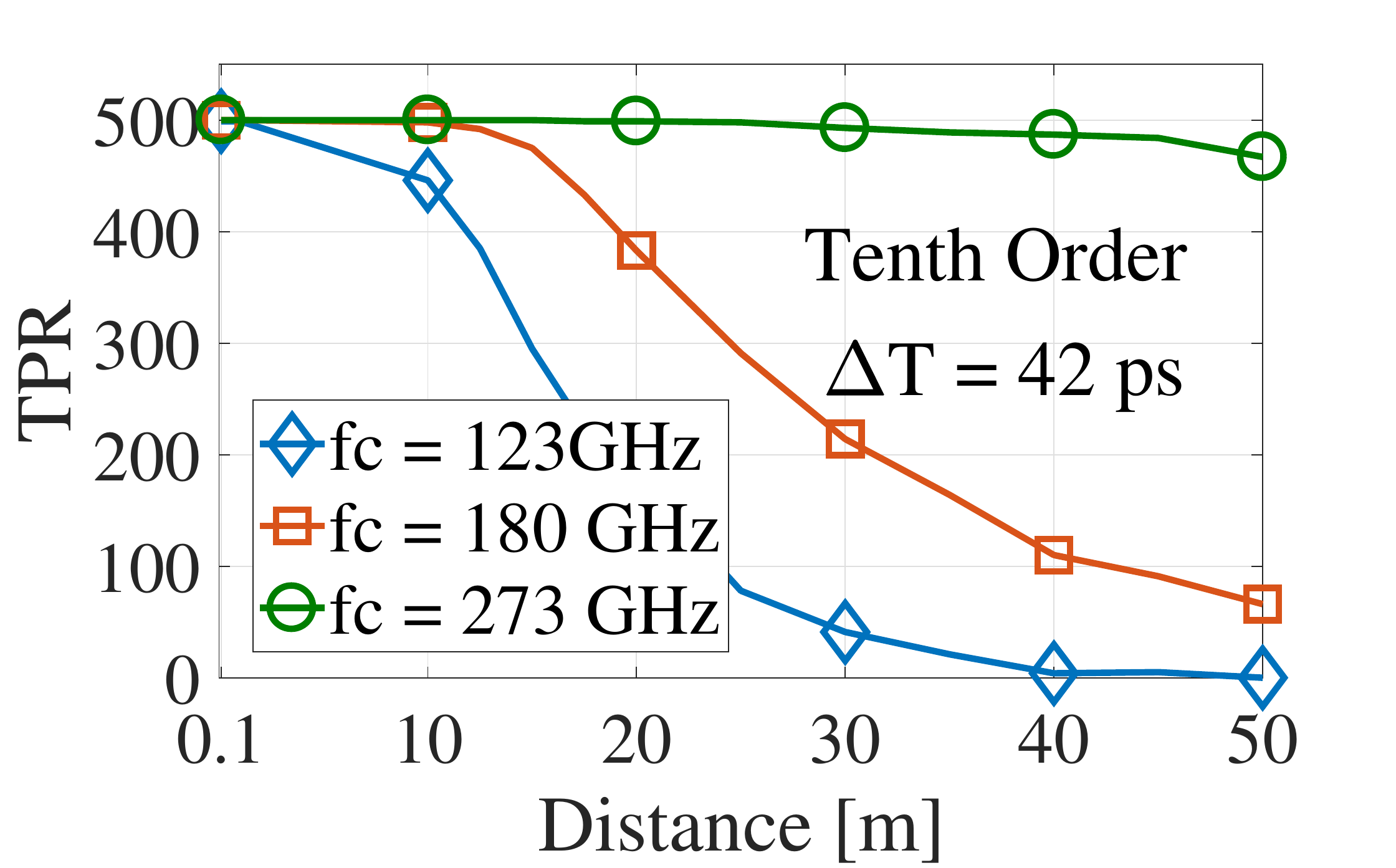}
		}
		\vspace{-0.75mm}
		\\
		\subfigure{
			\includegraphics[width=0.425\columnwidth,height=2.19cm]{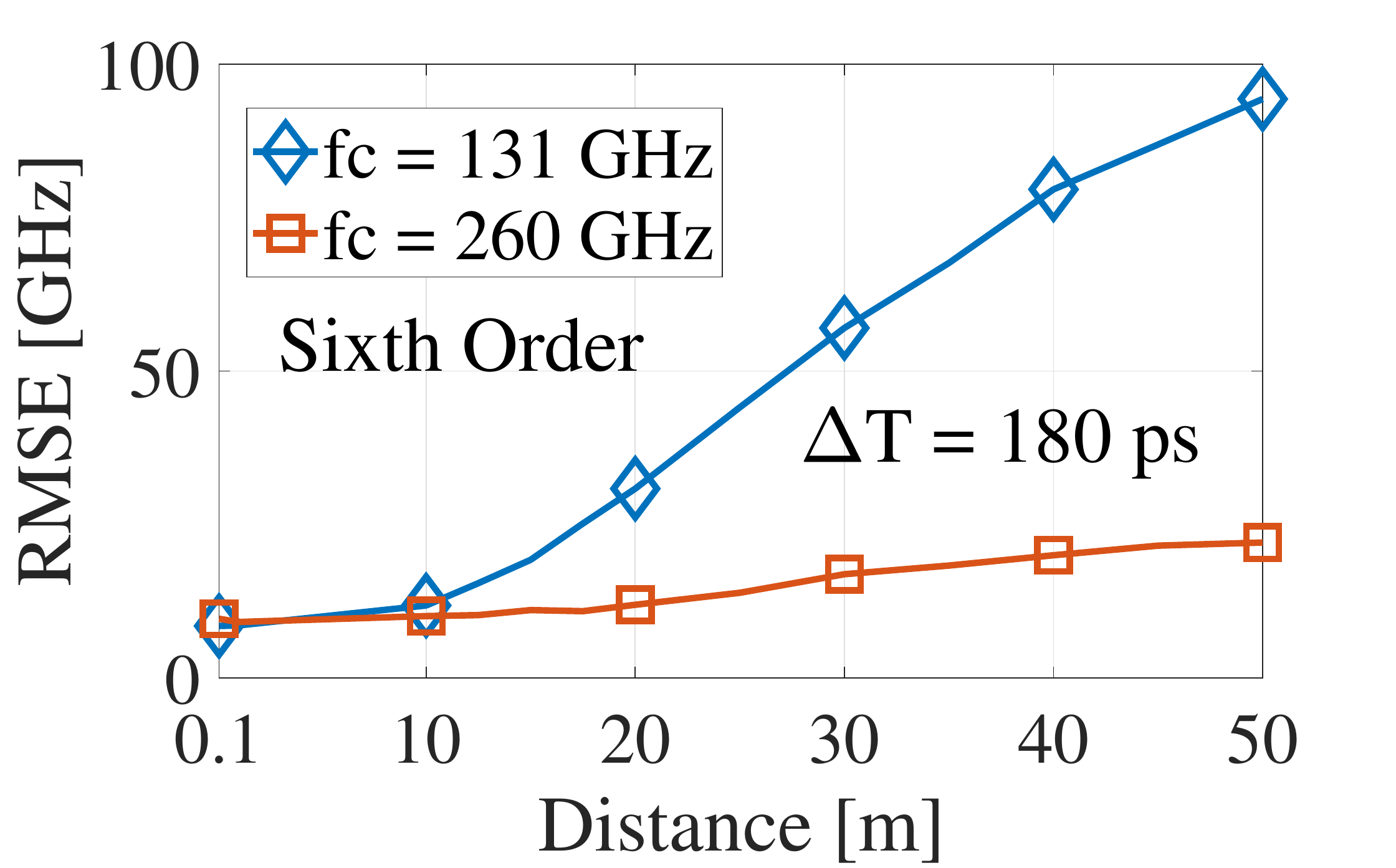}
		}
		\hspace{-5mm}
		\vspace{-0.75mm}
		\subfigure{
			\includegraphics[width=0.425\columnwidth,height=2.19cm]{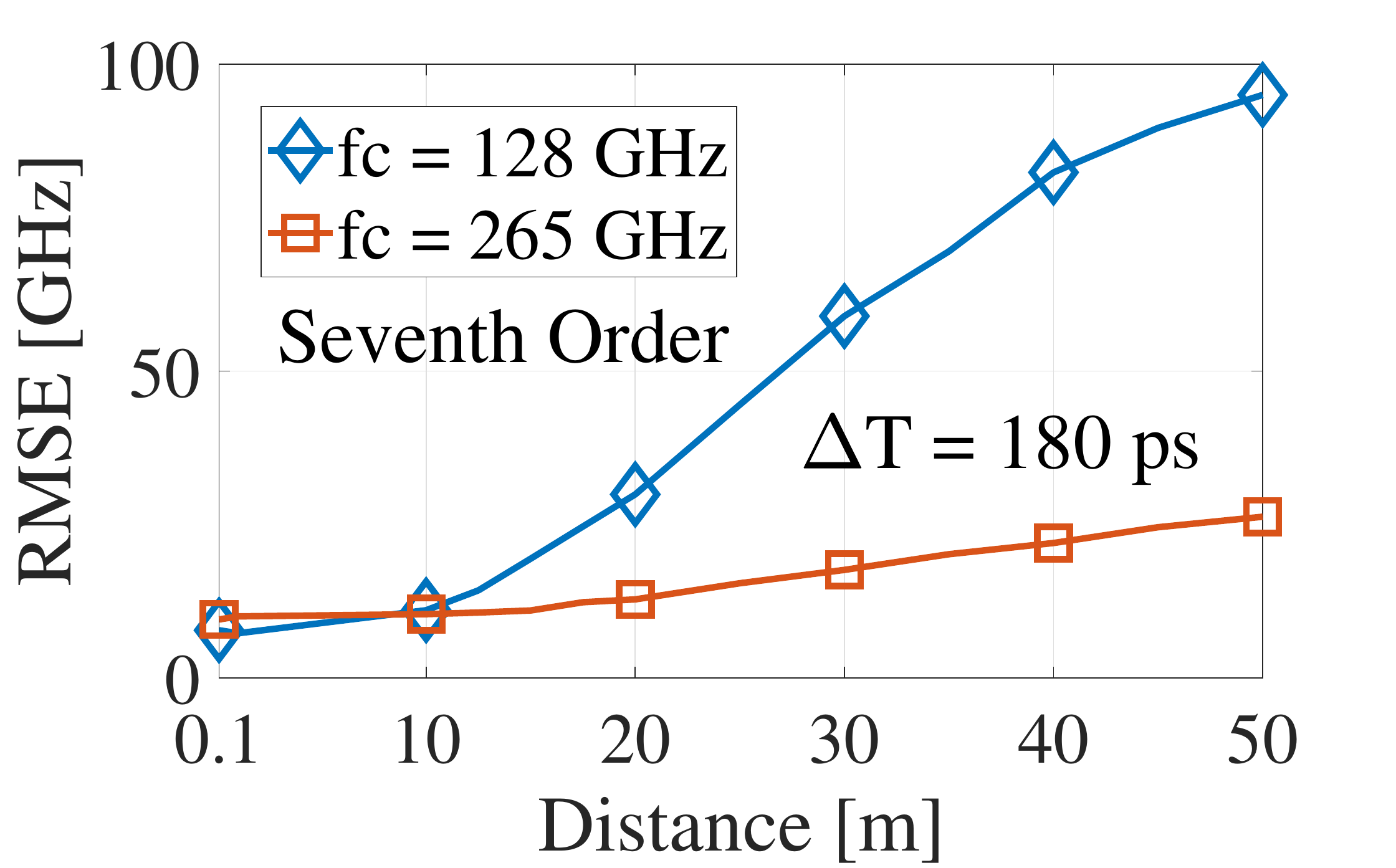}
		}
		\hspace{-5mm}
		\vspace{-0.75mm}
		\subfigure{
			\includegraphics[width=0.425\columnwidth,height=2.19cm]{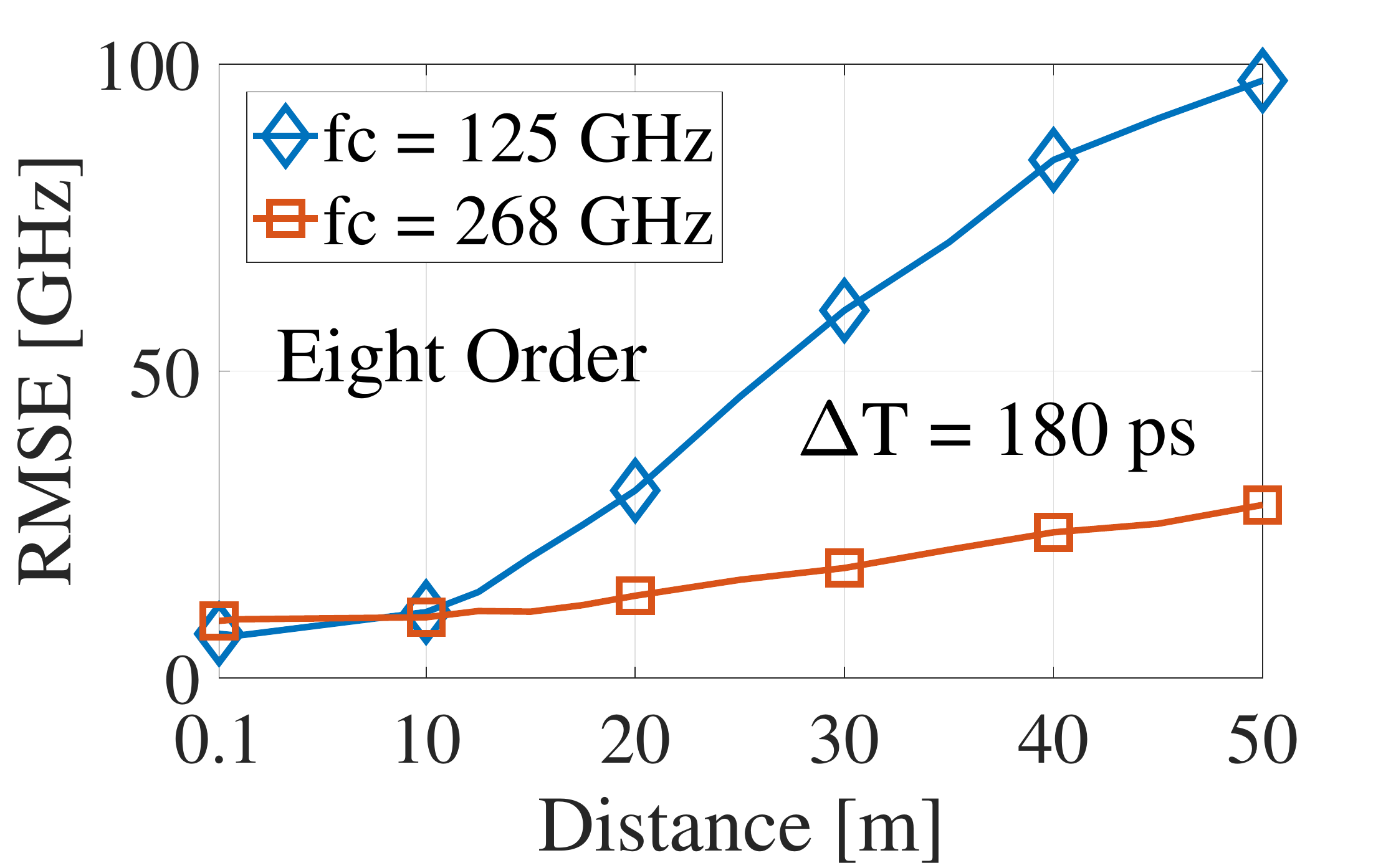}
		}
		\hspace{-5mm}
		\vspace{-0.75mm}
		\subfigure{
			\includegraphics[width=0.425\columnwidth,height=2.19cm]{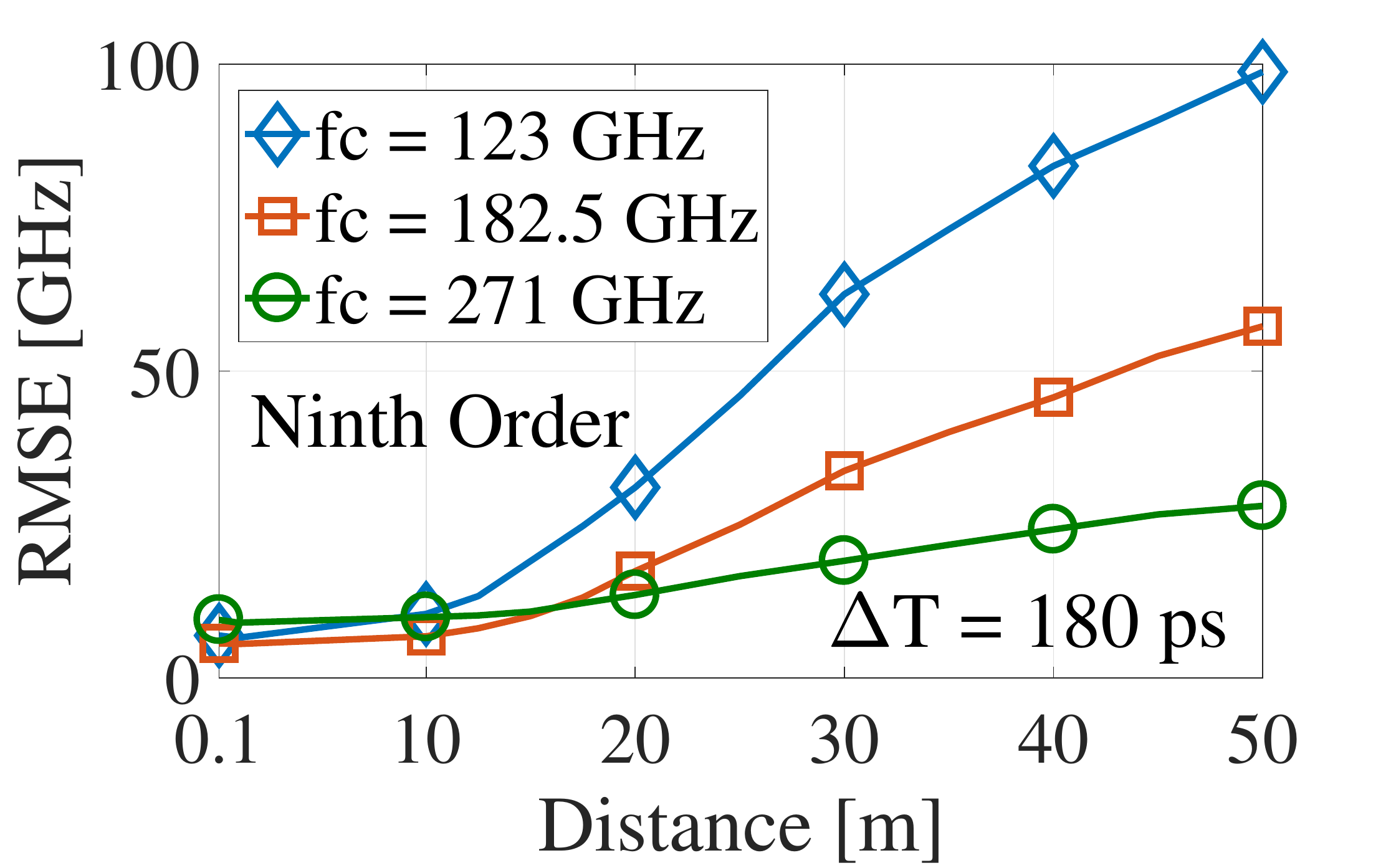}
		}
		\hspace{-5mm}
		\vspace{-0.75mm}
		\subfigure{
			\includegraphics[width=0.425\columnwidth,height=2.19cm]{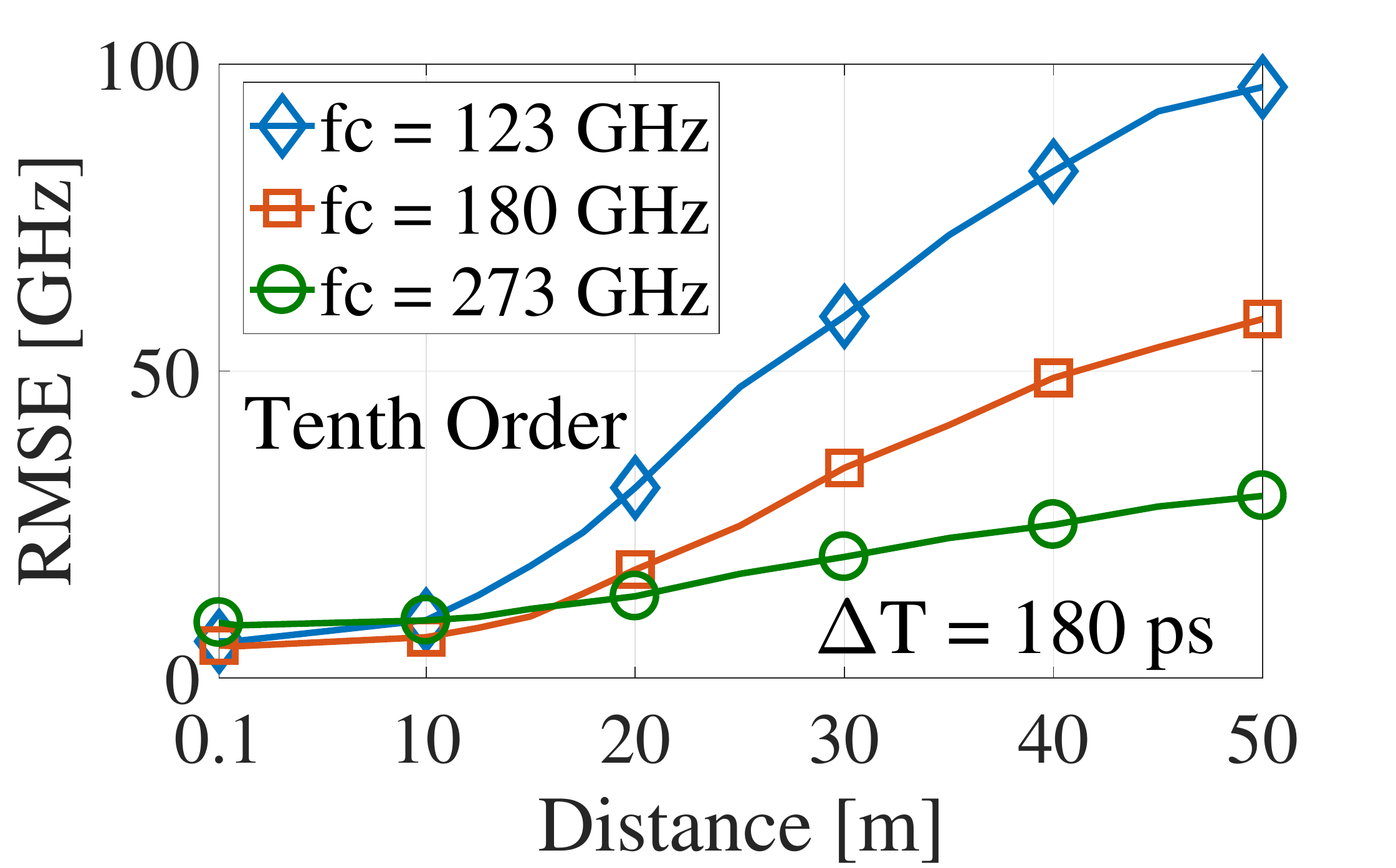}
		}
		\vspace{-0.75mm}
		\\
		\subfigure{
			\includegraphics[width=0.425\columnwidth,height=2.19cm]{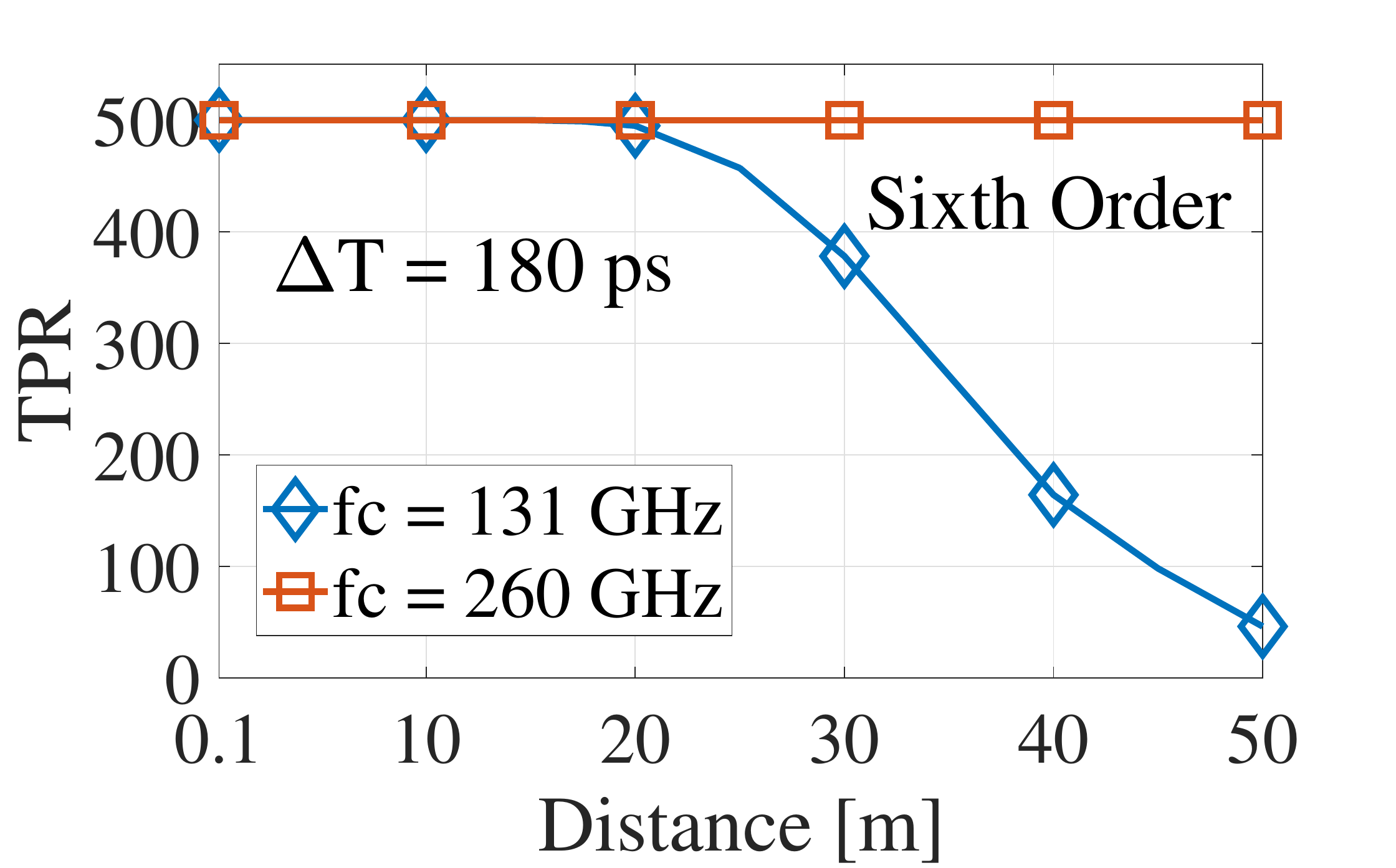}
		}
		\hspace{-5mm}
		\subfigure{
			\includegraphics[width=0.425\columnwidth,height=2.19cm]{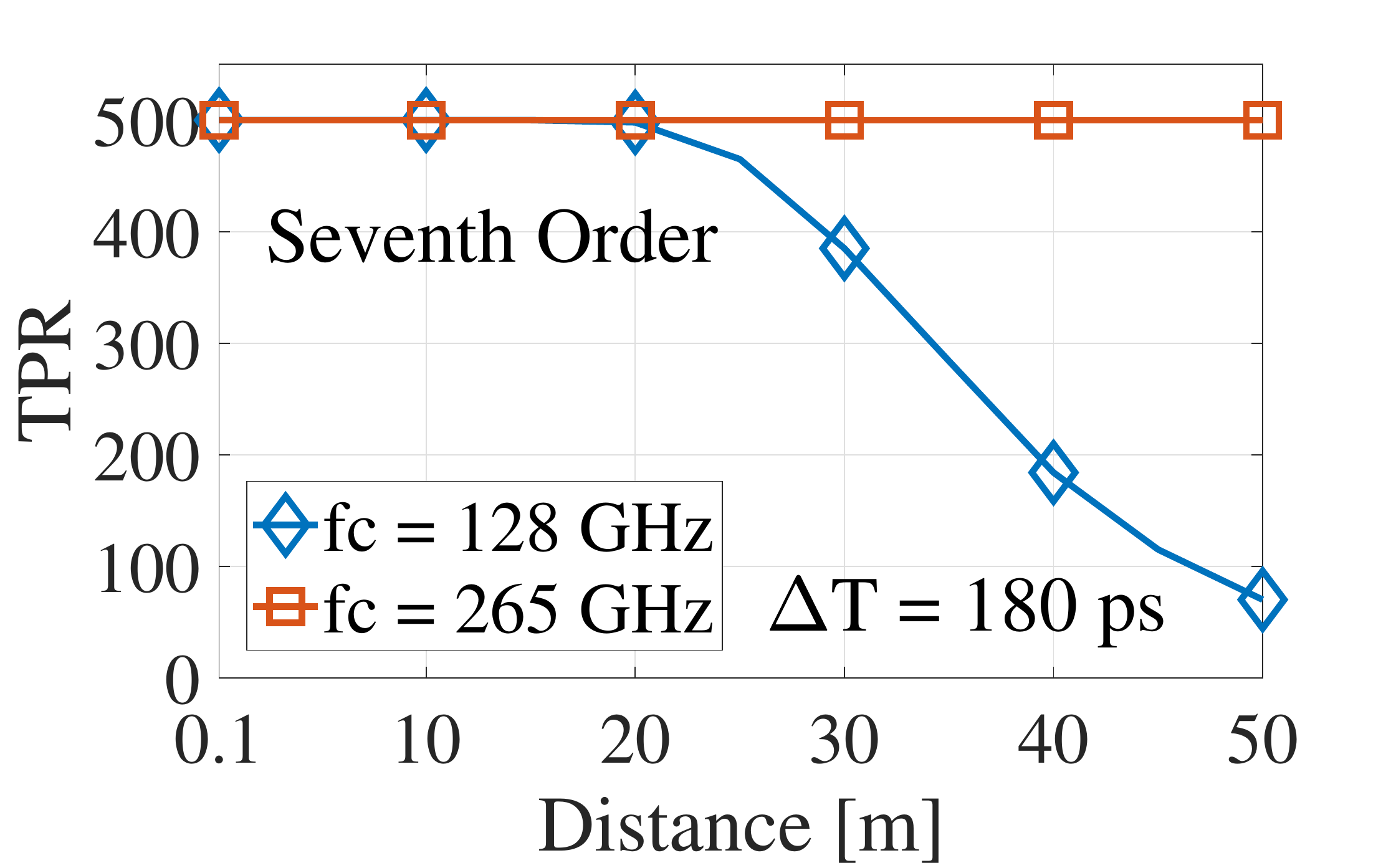}
		}
		\hspace{-5mm}
		\subfigure{
			\includegraphics[width=0.425\columnwidth,height=2.19cm]{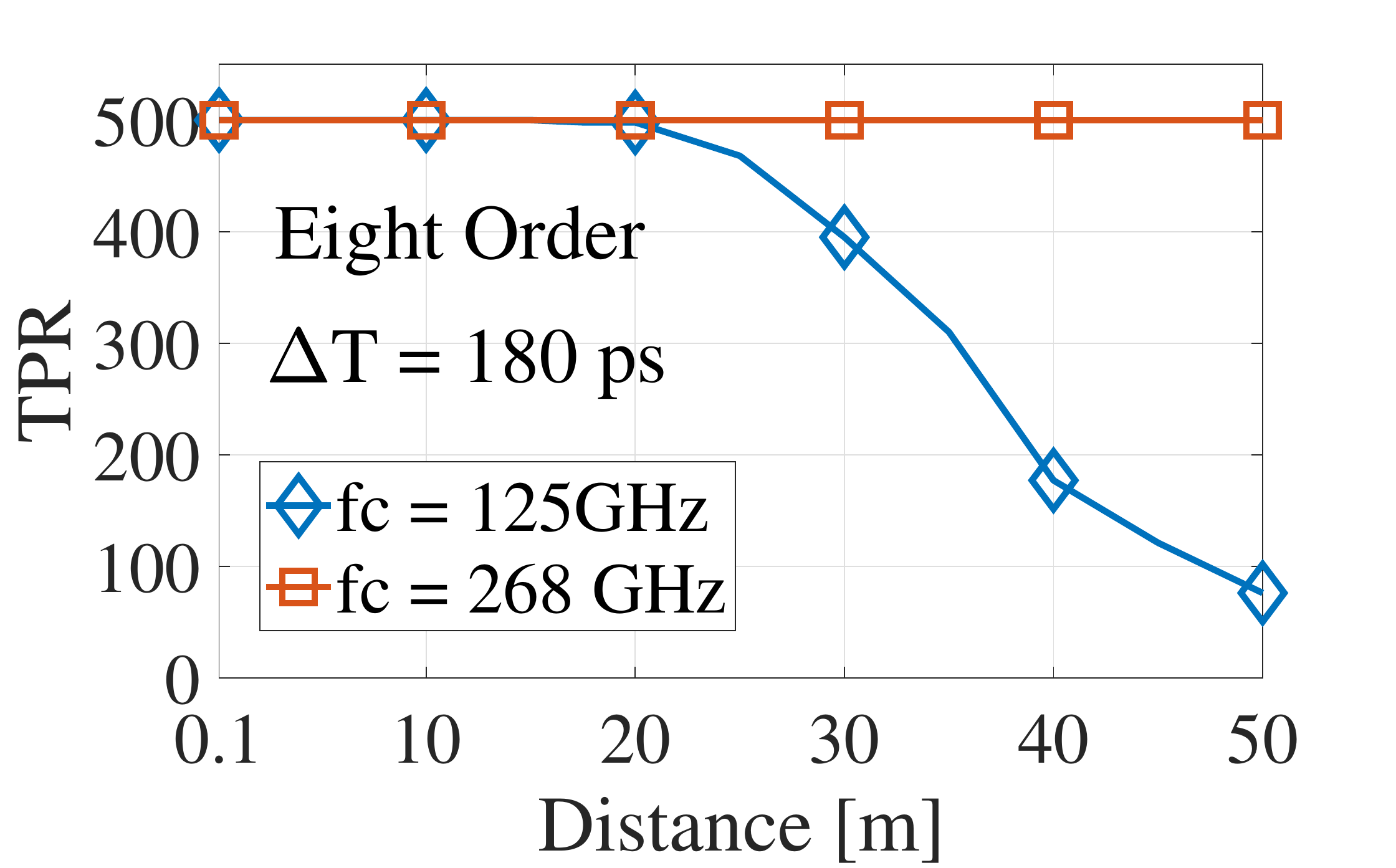}
		}
		\hspace{-5mm}
		\subfigure{
			\includegraphics[width=0.425\columnwidth,height=2.19cm]{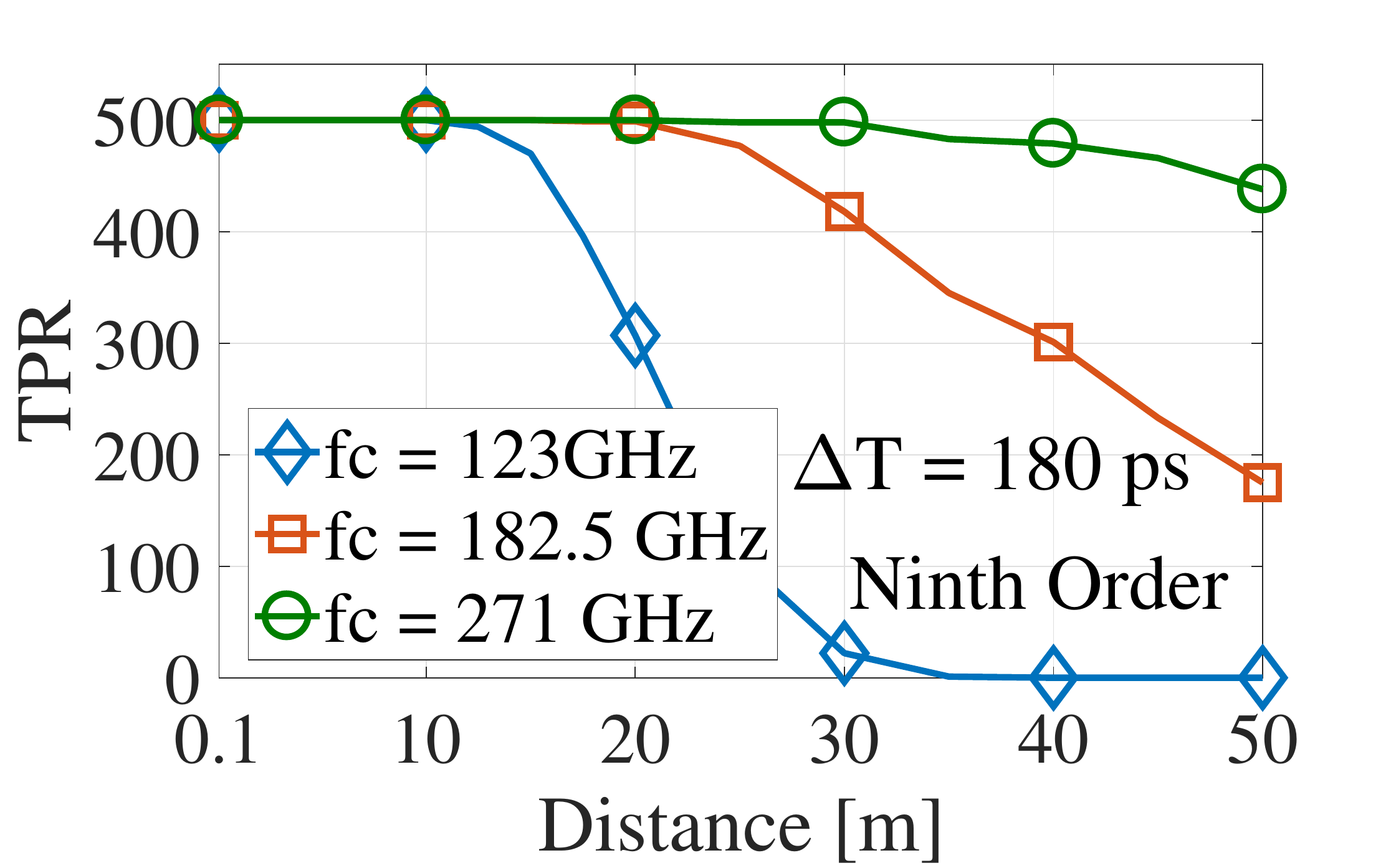}
		}
		\hspace{-5mm}
		\subfigure{
			\includegraphics[width=0.425\columnwidth,height=2.19cm]{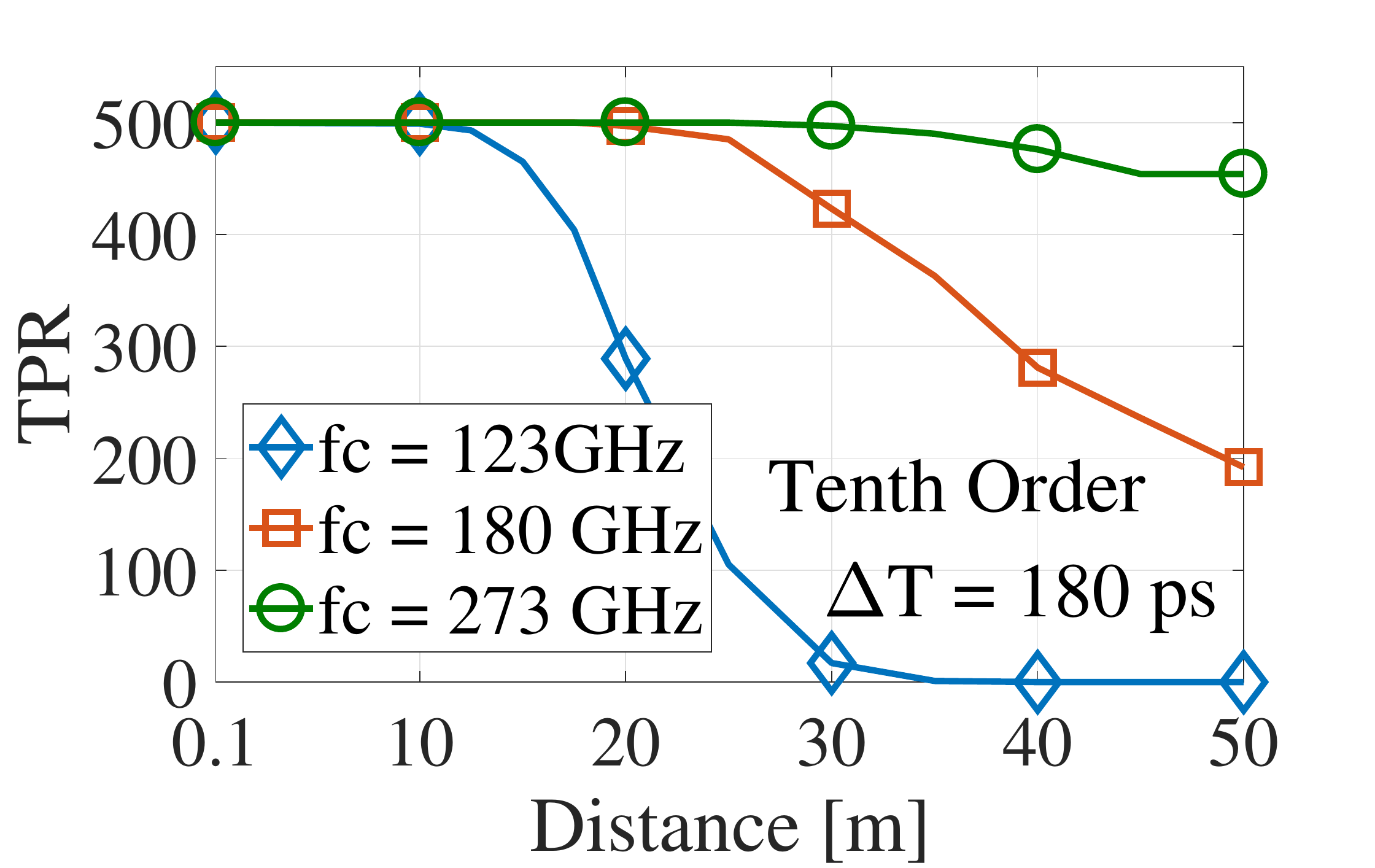}
		}
		\vspace{-5mm}
		\caption{Center frequency estimation and classification accuracy for higher order Gaussian pulses }
		\label{fig:FC_RMSE}
		\vspace{-5mm}
	\end{figure*}

Fig. \ref{fig:FC_RMSE} shows the RMSE and TPR as a function of path length for center frequency estimation and classification respectively. We make the following observations: \textbf{1)} The center frequency estimation and classification accuracy decrease rapidly with an increase in path length for Gaussian pulses with lower center frequencies. A possible explanation for this outcome is because the Gaussian pulse of lower center frequency will have lesser half power bandwidth. Hence, a fewer number of frequency bins with significant p.s.d. values will be available for estimating the center frequency. \textbf{2)} The center frequency estimation and classification accuracy is better for larger snapshot observation time. A possible explanation for this outcome is due to the improved p.s.d. estimate provided by a large number of frequency bins. With snapshot observation time 180 ps, 100\% average TPR is achieved at path lengths 20 m and 10 m for $n=6\;\text{to}\;8 $,  and $n=9\;\text{and}\;10 $ order Gaussian pulses respectively. Whereas for  snapshot observation time 42 ps, 100\% TPR is achieved at a smaller path lengths of 10 m and 1 m for $n=6\;\text{to}\;8 $,  and $n=9\;\text{and}\;10 $ order Gaussian pulses respectively. For $\Delta T = 42 \;\si{\pico}s$ and at a path length of 15 m, the confusion matrix in Table \ref{tabl:Table2} captures the fact that, the classification accuracy increases with increase in center frequency and lower center frequencies are erroneously classified only as higher center frequencies.
	\begin{table}
		\centering
		\caption{$E_{g}$ for Gaussian pulses with $P$ = 0.1 \si{\micro}W.}
		\label{tabl:ERG}
		\vspace{-4mm}
		\resizebox{0.7\columnwidth}{0.1\columnwidth}{\begin{tabular}{|c|c|c|c|c|c|}
				\hline
				$n$              &$f_{c}$ [GHz]  & $E_{g}$ [aJ]   & 	$n$                 & $f_{c}$ [GHz]   & $E_{g}$ [aJ]   \\ \hline
				\multirow{2}{*}{6} & 131 & 2.97 & \multirow{3}{*}{9}  & 123   & 3.88 \\ \cline{2-3} \cline{5-6} 
				& 260 & 1.49 &                     & 182.5 & 2.61 \\ \cline{1-3} \cline{5-6} 
				\multirow{2}{*}{7} & 128 & 3.28 &                     & 271   & 1.76 \\ \cline{2-6} 
				& 265 & 1.58 & \multirow{3}{*}{10} & 123   & 4.09 \\ \cline{1-3} \cline{5-6} 
				\multirow{2}{*}{8} & 125 & 3.60 &                     & 180   & 2.79 \\ \cline{2-3} \cline{5-6} 
				& 268 & 1.69 &                     & 273   & 1.84 \\ \hline
		\end{tabular}}
		\vspace{-4.5mm}
	\end{table}
	\vspace{-2.5mm}
	\begin{table}
		\centering
		\caption{Confusion Matrix for center frequency classification of  $9^{\text{th}}$ and $10^{\text{th}}$ order Gaussian pulse. The path length is 15 m and $\Delta T = 42\;\si{\pico}s$}
		\vspace{-4.25mm}
		\label{tabl:Table2}
		\resizebox{\columnwidth}{0.20\columnwidth}{\begin{tabular}{|c|c|c|c|c|c|c|c|}
				\hline
				\multicolumn{2}{|c|}{\multirow{3}{*}{}}                     & \multicolumn{3}{c|}{Ninth Order}                                   & \multicolumn{3}{c|}{Tenth Order}                                   \\ \cline{3-8} 
				\multicolumn{2}{|c|}{}                                      & \multicolumn{6}{c|}{True Frequency}                                                                                                     \\ \cline{3-8} 
				\multicolumn{2}{|c|}{}                                      & $f_{c1}$                 & $f_{c2}$                  & $f_{c3}$                  & $f_{c1}$                  & $f_{c2}$                  & $f_{c3}$                  \\ \hline
				\multirow{6}{*}{\rotatebox{90}{\parbox[c]{1cm}{Estimated\\Frequency}}} & \multirow{2}{*}{$f_{c1}$} & \multirow{2}{*}{\textbf{308}} & \multirow{2}{*}{0}   & \multirow{2}{*}{0}   & \multirow{2}{*}{\textbf{304}} & \multirow{2}{*}{0}   & \multirow{2}{*}{0}   \\
				&                      &                      &                      &                      &                      &                      &                      \\ \cline{2-8} 
				& \multirow{2}{*}{$f_{c2}$} & \multirow{2}{*}{180} & \multirow{2}{*}{\textbf{468}} & \multirow{2}{*}{0}   & \multirow{2}{*}{188} & \multirow{2}{*}{\textbf{462}} & \multirow{2}{*}{0}   \\
				&                      &                      &                      &                      &                      &                      &                      \\ \cline{2-8} 
				& \multirow{2}{*}{$f_{c3}$} & \multirow{2}{*}{12}  & \multirow{2}{*}{32}  & \multirow{2}{*}{\textbf{500}} & \multirow{2}{*}{8}   & \multirow{2}{*}{38}  & \multirow{2}{*}{\textbf{500}} \\
				&                      &                      &                      &                      &                      &                      &                      \\ \hline
				\multicolumn{2}{|c|}{TPR}                                   & 308                  & 468                  & 500                  & 307                  & 462                  & 500                  \\ \hline
				\multicolumn{2}{|c|}{Over all TPR}                          & \multicolumn{3}{c|}{\textbf{425.33}}                                        & \multicolumn{3}{c|}{\textbf{422}}                                           \\ \hline
		\end{tabular}}
		\vspace{-6mm}
	\end{table}
\vspace{-4mm}
	\vspace{-2mm}
	\section{Conclusion}
	\vspace{-0.5mm}
	We have proposed a framework to simultaneously localize and estimate center frequency using a single higher order Gaussian mmWave pulse. The performance of the proposed framework is evaluated for graphene based transceivers operating in 100-325 GHz mmWave band. Our investigation shows that for a given pulse power, Gaussian pulses with lower center frequencies provide better DOA estimation accuracy whereas good center frequency classification accuracy is provided by Gaussian pulses with higher center frequency. Further, simulation results show that, for snapshot observation interval of 180 ps, it is possible to achieve DOA estimation accuracy below $1^{\circ}$ and classify three different center frequencies with 100\% accuracy for path lengths up to 20 m and 10 m respectively.  In future, event classification will be investigated based on the order of Gaussian pulses.
	\vspace{-2mm}
	\section*{Acknowledgment}
	\vspace{-0.5mm}
	The proposed work is supported by SERB, GOI under order no. SB/S3/EECE/210/2016
	\vspace{-5mm}
	\bibliographystyle{ACM-Reference-Format}
	\bibliography{MOBICOM_IOT}
\end{document}